\documentclass[review,numbers,sort&compress,3p,times,12pt]{elsarticle}
\usepackage{multirow,setspace,times,amssymb,amsmath,graphicx,color,rotating,subfigure,url}
\usepackage{lineno,color}
\usepackage{natbib}
\usepackage{booktabs}
\usepackage{longtable}
\usepackage{rotating}
\usepackage{lscape}
\usepackage{threeparttable}
\usepackage{bm}
\graphicspath{{Figures/}}
\usepackage[table]{xcolor}
\usepackage{tabularx}
\usepackage{graphicx} 
\usepackage{epstopdf}
\usepackage{mathrsfs}
\usepackage{makecell}
\usepackage[bookmarks=true,colorlinks,linkcolor=blue,anchorcolor=blue,citecolor=blue,unicode,urlcolor=blue]{hyperref}
\usepackage{bookmark}
\usepackage{todonotes}
\usepackage{ragged2e}
\usepackage[font=small]{caption}
\usepackage{amsthm}
\newtheorem{theorem}{Theorem}
\newtheorem{lemma}{Lemma}
\usepackage{algorithm}
\usepackage{algpseudocode}
\usepackage{lineno}
\usepackage{dcolumn}
\hypersetup{CJKbookmarks=true}%
\journal{}
\begin{document}
\captionsetup[figure]{font={bf},name={Fig.},labelsep=period}
\captionsetup[table]{labelfont={bf},name={Table},labelsep=newline,singlelinecheck=false}

\begin{frontmatter}

\title{On the existence of Ulanowicz's optimal structural resilience in complex networks}	
\author[SB,RCE]{Si-Yao Wei}
\author[SB,RCE,Math]{Wei-Xing Zhou\corref{cor1}}
\ead{wxzhou@ecust.edu.cn} 
\cortext[cor1]{Corresponding author.}
\address[SB]{School of Business, East China University of Science and Technology, Shanghai 200237, China}
\address[RCE]{Research Center for Econophysics, East China University of Science and Technology, Shanghai 200237, China}
\address[Math]{School of Mathematics, East China University of Science and Technology, Shanghai 200237, China}

\begin{abstract}
This study provides a foundational theoretical investigation into the mathematical existence and asymptotic properties of Ulanowicz's structural resilience. While ecological evidence suggests that sustainable systems gravitate toward an optimal efficiency-redundancy balance at $\alpha = 1/\mathrm{e}$, the mathematical attainability of this configuration across broader network topologies remains unverified. We rigorously prove that while optimal resilience is structurally unattainable in two-node networks, there exists at least one optimal flow configuration within the feasible probability space for any weighted and directed network with the network size $N_\mathcal{V} \geq 3$ and no self-loops. To make the derivations analytically tractable, we introduce a parameterized symmetric network model with uniform marginal distributions. Using this stylized ansatz, our analytical and numerical results reveal that maintaining the optimal state requires distinct asymptotic scaling behaviors as $N_\mathcal{V}$ increases: adjacent primary links scale as $O(N_\mathcal{V}^{-1})$, whereas non-adjacent background links exhibit a steeper quadratic decay of $O(N_\mathcal{V}^{-2})$ with specific logarithmic corrections. Rather than serving as an immediate engineering tool, this work establishes a rigorous mathematical boundary for the optimal resilience framework, demonstrating analytically how an optimally resilient system differentiates into high-throughput primary channels and sparse redundancy pathways.
\end{abstract}

\end{frontmatter}

\section{Introduction}

Resilience is a fundamental property of networked complex systems \citep{FJ-Liu-Li-Ma-Szymanski-Stanley-Gao-2022-PhysRep}. Consequently, the definition and measurement of resilience in complex networks have emerged as focal points of interdisciplinary research, spanning from ecological stability \citep{FJ-Liu-Yang-Chen-Zhang-Zhang-Zhao-Jiang-2009-CommunNonlinearSciNumerSimul,FJ-Seidl-Spies-Peterson-Stephens-Hicke-2016-JApplEcol} and infrastructure reliability \citep{FJ-Chen-Ma-Chen-Yang-2024-TranspResPartD} to the robustness of global economic systems \citep{FJ-Brunnermeier-2024-JFinanc} and supply chains \citep{FJ-Cohen-Cui-Doetsch-Ernst-Huchzermeier-Kouvelis-Lee-Matsuo-Tsay-2022-JOperManag,FJ-Ivanov-2024-Omega-IntJManageSci}. Various studies suggest that resilient complex systems must simultaneously achieve high operational efficiency and sufficient redundancy to withstand disruptions \citep{FJ-Lucker-TimoninaFarkas-Seifert-2025-ProdOperManag,FJ-Kahiluoto-Makinen-Kaseva-2020-IntJOperProdManage,FJ-Yang-Wu-Sun-Zhang-2024-JRiskRes}. However, these two objectives are often conflicting: highly efficient networks tend to be fragile, whereas overly redundant systems incur excessive operational costs. Understanding and quantifying this trade-off has therefore become a fundamental research challenge in network optimization and robust system design \citep{FJ-Pettit-Croxton-Fiksel-2019-JBusLogist,FJ-Zhu-Bao-Qin-Sun-Shia-Chen-2025-AnnOperRes}.

Entropy-based measures have recently attracted increasing attention as tools for characterizing structural robustness in complex networks. In the field of theoretical ecology, Ulanowicz proposed an information-theoretic framework to quantify this tension, defining system fitness or structural resilience as a function of the balance between efficiency and redundancy \citep{FJ-Ulanowicz-1979-Oecologia,FJ-Ulanowicz-Norden-1990-IntJSystSci,FJ-Ulanowicz-2009-EcolModel}. Within this framework, efficiency is associated with the mutual information of the flow distribution, while redundancy corresponds to the conditional entropy. Central to this theory is the ``{\it Window of Vitality}'' hypothesis, which suggests that sustainable and resilient ecosystems do not maximize efficiency or redundancy. Instead, they gravitate toward an optimal configuration where the degree of order, denoted by $\alpha$, is approximately $1/\mathrm{e}\approx 0.3679$. At this critical point, the system is argued to possess sufficient articulation to function effectively and enough diversity to adapt to changing environments.

Despite the empirical success of Ulanowicz's metrics in evaluating biological food webs \citep{FJ-Ulanowicz-2009-EcolModel}, industrial metabolic networks \citep{FJ-Liang-Yu-Kharrazi-Fath-Feng-Daigger-Chen-Ma-Zhu-Mi-Yang-2020-NatFood,FJ-Luo-Yu-Kharrazi-Fath-Matsubae-Liang-Chen-Zhu-Ma-Hu-2024-NatFood,FJ-Xia-Fu-Ke-Wang-Liang-Yang-2025-ApplEnergy}, and international trade networks \citep{FJ-Kharrazi-Rovenskaya-Fath-Yarime-Kraines-2013-EcolEcon}, fundamental theoretical questions remain unanswered. First, while evolutionary selection drives ecological systems toward this optimal $1/\mathrm{e}$ configuration, it remains unclear whether this optimal state is mathematically attainable within the feasible probability space of directed flow networks. Second, if such an optimal state exists, how do network size and structural topology dictate the scaling of link weights required to maintain it? Because existing literature predominantly focuses on specific, data-driven empirical instances, there remains a critical gap in the rigorous mathematical formulation of these entropy-based measures.

In summary, the primary objective of this paper is to establish a rigorous mathematical foundation for Ulanowicz's optimal structural resilience by clarifying its existence and asymptotic properties within an information-theoretic framework. Acknowledging the immense complexity of real-world heterogeneous systems, our work is intended as a foundational theoretical clarification rather than an immediate empirical application. Specifically, our main contributions are three-fold: (1) General existence proofs. We rigorously prove that while optimal resilience ($R=1/\mathrm{e}$) is structurally unattainable in two-node systems, there exists at least one optimal flow configuration within the feasible probability space for any weighted and directed network with $N_\mathcal{V} \ge 3$ and no self-loops. (2) Stylized symmetric construction. To make the analytical derivation of this optimal state mathematically tractable, we introduce a stylized, highly symmetric network model parameterized by three link types $(x,y,z)$. While this symmetric ansatz narrows the scope of the topology, it provides an exact framework to observe the mechanics of the trade-off between efficiency and redundancy. (3) Asymptotic scaling analysis. Using this symmetric model, we derive explicit asymptotic scaling laws for the link weights as $N_\mathcal{V} \to \infty$. These derivations serve as a specialized mathematical demonstration of how topological weights must scale to maintain the $1/\mathrm{e}$ optimal state as network size increases. By delineating these mathematical boundaries, we hope to provide a theoretical benchmark for future studies exploring optimal resilience in more complex, heterogeneous, and dynamically constrained networks.

The remainder of this paper is organized as follows. Section~\ref{Section_Definitions} formalizes the definitions of efficiency, redundancy, and structural resilience. Section~\ref{Section_Existence} presents the proofs of the existence theorems, while Section~\ref{Section_Results} analyzes the $N_\mathcal{V}$-node symmetric model, including its asymptotic scaling laws. Finally, Section~\ref{Section_Conclusion} concludes and discusses the limitations.

\section{Definitions}
\label{Section_Definitions}

In this section, we introduce metrics in Ulanowicz's entropy-based resilience framework, including efficiency $e$, redundancy $r$, ratio of order $\alpha$, and resilience $R$. Note that this framework was fundamentally designed for directed flow networks. Consequently, all network models in this study are weighted, directed, and constructed without self-loops.

\subsection{Efficiency and redundancy}

Consider a digraph $\mathcal{G}=\langle\mathcal{V},\mathcal{E}\rangle$, where $\mathcal{V}$ represents the vertex set with $N_\mathcal{V}$ vertices and $\mathcal{E}$ represents the link set with $N_\mathcal{E}$ links. Let $f_{ij}$ denote the flow from vertex $i$ to vertex $j$. The out-strength of vertex $i$ is $s_i^{\mathrm{out}}=\sum\limits_{j=1}^{N_\mathcal{V}}f_{ij}$ and the in-strength of vertex $j$ is $s_j^{\mathrm{in}}=\sum\limits_{i=1}^{N_\mathcal{V}}f_{ij}$. The throughput $s=\sum\limits_{i=1}^{N_\mathcal{V}}s_i^{\mathrm{out}}=\sum\limits_{j=1}^{N_\mathcal{V}}s_j^{\mathrm{in}}=\sum\limits_{i=1}^{N_\mathcal{V}}\sum\limits_{j=1}^{N_\mathcal{V}}f_{ij}$. 

In information theory, entropy measures a system's uncertainty (or diversity). A higher entropy reflects greater unpredictability and diversity, indicating an enhanced adaptability to changing conditions and an improved capacity of the system to absorb shocks \citep{FJ-Khakifirooz-Fathi-Dolgui-Pardalos-2025-IntJProdRes,FJ-Reggiani-2022-NetwSpatEcon}. Accordingly, we can adopt joint entropy to characterize a system's overall structural properties \citep{FJ-Rutledge-Basore-Mulholland-1976-JTheorBiol,FJ-Ulanowicz-1979-Oecologia}, which is also called as ``capacity'' for system development \citep{FJ-Ulanowicz-Norden-1990-IntJSystSci}:
\begin{equation}
    H = -\sum\limits_{i=1}^{N_\mathcal{V}}\sum\limits_{j=1}^{N_\mathcal{V}} p_{ij}\ln p_{ij}.
\end{equation}

Given that joint entropy is equal to the sum of mutual information and conditional entropy, we further introduce two underlying quantities, efficiency and redundancy, which demonstrate opposing properties of a system. Efficiency embodies the flow articulation within networked configurations, which tends to increase due to preferential interactions between nodes \citep{FJ-Kharrazi-Rovenskaya-Fath-2017-PloSOne}. In information theory, a higher mutual information indicates a increased reduction in uncertainty when some information is known and a stronger statistical inter-dependence between variables. Hence, we define the efficiency $e$ of digraph $\mathcal{G}$ as follows:
\begin{equation}
    e=\sum\limits_{i=1}^{N_\mathcal{V}}\sum\limits_{j=1}^{N_\mathcal{V}} p_{ij}\ln\frac{p_{ij}}{p_{i}^{\mathrm{out}}p_{j}^{\mathrm{in}}}
    =\sum_{i=1}^{N_\mathcal{V}}\sum_{j=1}^{N_\mathcal{V}}\frac{f_{ij}}{s} \ln\frac{f_{ij}s}{s_i^{\mathrm{out}}s_j^{\mathrm{in}}},
    \label{Eq_Def_Efficiency}
\end{equation}
where
\begin{equation}
    p_{ij}=\frac{f_{ij}}{s},~~ p_{i}^{\mathrm{out}}=\frac{s_i^{\mathrm{out}}}{s},~~ p_{j}^{\mathrm{in}}=\frac{s_j^{\mathrm{in}}}{s}.
\end{equation}

Redundancy embodies the diversity of pathways, which is critical for a system's capacity adapting to changing environmental conditions arising from shocks or disturbances \citep{FJ-Luo-Yu-Kharrazi-Fath-Matsubae-Liang-Chen-Zhu-Ma-Hu-2024-NatFood}. Note that our perspective departs from studies that focus primarily on overall uncertainty and instead emphasizes the trade-off between efficiency and redundancy. In this context, redundancy is conceptualized as the residual uncertainty conditional on partial information, which corresponds to the definition of conditional entropy in information theory. A higher conditional entropy indicates a greater availability of alternative pathways and a consequently higher system redundancy. Hence, the redundancy $r$ of digraph $\mathcal{G}$ can be defined as
\begin{equation}
r
=-\sum\limits_{i=1}^{N_\mathcal{V}}\sum\limits_{j=1}^{N_\mathcal{V}} p_{ij}\ln \frac{p_{ij}}{p_{i}^{\mathrm{out}}}-\sum\limits_{i=1}^{N_\mathcal{V}}\sum\limits_{j=1}^{N_\mathcal{V}} p_{ij}\ln\frac{p_{ij}}{p_{j}^{\mathrm{in}}}
=\sum_{i=1}^{N_\mathcal{V}}\sum_{j=1}^{N_\mathcal{V}}\frac{f_{ij}}{s}\ln\frac{s_i^{\mathrm{out}}s_j^{\mathrm{in}}}{f_{ij}^2}.
\label{Eq_Def_Redundancy}
\end{equation}

\subsection{Ulanowicz's structural resilience}

Based on these two system properties, the ratio $\alpha$, a more comprehensive metric to indicate the order of a system, is proposed for reflecting the trade-off between efficiency and redundancy \citep{FJ-Goerner-Lietaer-Ulanowicz-2009-EcolEcon,FJ-Ulanowicz-Goerner-Lietaer-Gomez-2009-EcolComplex,FJ-Ulanowicz-1979-Oecologia}, expressed as
\begin{equation}
    \alpha=\frac{e}{H}=\frac{e}{e+r},
    \label{Eq_Def_Alpha}
\end{equation}
where $0\leq\alpha\leq1$. 

\begin{figure}[htb!]
    \centering
    \includegraphics[width=1\linewidth]{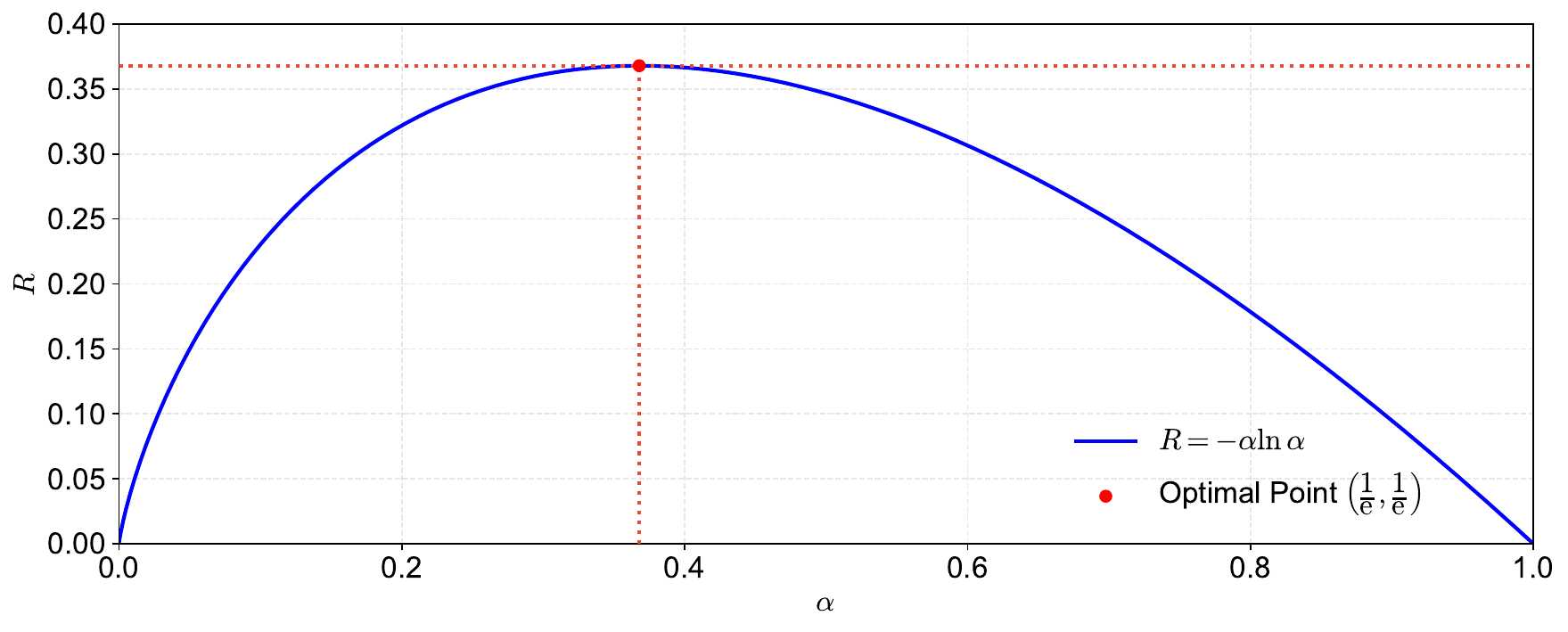}
    \caption{The function $R=-\alpha\ln\alpha$.}
    \label{Fig_alpha_log_alpha}
\end{figure}

Inspired by ecological systems, whose order parameters are often observed to be close to $1/{\mathrm{e}}$ \citep{FJ-Zorach-Ulanowicz-2003-Complexity}, Ulanowicz defined the fitness $F$ of a system for change to be the product of $\alpha$ and the Boltzmann measure of its disorder such that $F=-c\alpha\ln\alpha$, where $c$ is an appropriate scalar constant and $\mathrm{e}$ is Euler's number \citep{FJ-Ulanowicz-2009-EcolModel}. At $\alpha=1/{\mathrm{e}}$, the first derivative $F'=0$, indicating that the system's fitness is maximized (or optimal). The underlying assumption is that ecosystems exhibit superior trade-off because they have undergone long-term natural selection \citep{FJ-Ulanowicz-2009-EcolModel,FJ-Liang-Yu-Kharrazi-Fath-Feng-Daigger-Chen-Ma-Zhu-Mi-Yang-2020-NatFood,FJ-Kharrazi-Rovenskaya-Fath-Yarime-Kraines-2013-EcolEcon}. Finally, let $c=1$, the resilience $R$ of $\mathcal{G}$ can be defined as
\begin{equation}
    R=-\alpha\ln\alpha,
    \label{Eq_Def_Resilience}
\end{equation}
with the function shown in Fig.~\ref{Fig_alpha_log_alpha}. When $\alpha>1/\mathrm{e}$, the system is more efficient and productive but more vulnerable, and vice versa it is more redundant but more inefficient. $R$ vanishes when $\alpha=0$ (overly redundant) or $\alpha=1$ (overly efficient).

\section{Existence of optimal resilience in complex networks}
\label{Section_Existence}

While empirical studies observe $\alpha$ nears $1/\mathrm{e}$ in ecological systems, it remains theoretically unclear whether this optimal state is a mathematically reachable configuration for other networks. Proving the attainability of $\alpha = 1/\mathrm{e}$ is a fundamental theoretical prerequisite. If intrinsic structural constraints (e.g., probability normalization and topological limitations) inherently prohibit a network from achieving this precise balance, then it cannot serve as a valid and computable objective for network optimization. Therefore, establishing its mathematical existence bounds the applicability of Ulanowicz's framework, translating it from an ecological observation into a rigorous prerequisite for robust network design, presented as follows.

\subsection{Two-node networks}

\begin{theorem}
\label{Thm_TwoNodes}
The optimal resilience ($R=1/\mathrm{e}$) does not exist in any two-node weighted and directed network without self-loops.
\end{theorem}

\begin{proof}
Consider a digraph $\mathcal{G}$ with two vertices. Let $p_{12} = p$ and $p_{21} = 1-p$, with $p_{11} = 0$ and $p_{22} = 0$. Also, we have $p_1^{\mathrm{out}} = p_2^{\mathrm{in}} = p_{12} = p$ and $p_2^{\mathrm{out}} = p_1^{\mathrm{in}} = p_{21} = 1-p$. Substituting them into Eq.~(\ref{Eq_Def_Efficiency}) yields
\begin{equation}
\begin{aligned}
e(p) &= \sum_{i=1}^{2}\sum_{j=1}^{2} p_{ij} \ln \frac{p_{ij}}{p_{i}^{\mathrm{out}} p_{j}^{\mathrm{in}}} \\
&= -p \ln p - (1-p) \ln (1-p).
\end{aligned}
\end{equation}
According to Eq.~(\ref{Eq_Def_Redundancy}), we have
\begin{equation}
    \begin{aligned}
r(p) &= -\sum_{i=1}^{2}\sum_{j=1}^{2} p_{ij} \ln \frac{p_{ij}}{p_{i}^{\mathrm{out}}} - \sum_{i=1}^{2}\sum_{j=1}^{2} p_{ij} \ln \frac{p_{ij}}{p_{j}^{\mathrm{in}}} \\
&= 0.
\end{aligned}
\end{equation}
Accordingly, the ratio \begin{equation}
    \alpha(p) = \frac{e(p)}{e(p) + r(p)}
    \equiv 1 \quad \text{for} \quad p \in [0,1]
\end{equation}
and the resilience
\begin{equation}
    R(p) = -\alpha(p)\ln\alpha(p) = -1 \cdot \ln(1) = 0,
\end{equation}
representing that a two-node weighted and directed network with no self-loops is deterministic and overly-efficient, and thereby cannot attain the optimal resilience ($R=1/\mathrm{e}$).
\end{proof}

\subsection{Networks with at least three nodes}

\begin{theorem}
\label{Thm_AtleastThreeNodes}
For any weighted and directed network with at least three nodes and no self-loops, there exists at least one joint distribution such that its resilience is optimal (i.e., $R=1/\mathrm{e}$).
\end{theorem}

\begin{proof}
Let ${N_\mathcal{V}}\geq 3$. 
Define the set of all feasible network configurations as
\begin{equation}
    \mathcal{P}=\left\{{\bm{p}}=(p_{ij})_{i,j=1}^{N_\mathcal{V}}: p_{ij}\ge 0,\; p_{ii}=0,\; \sum_{i=1}^{N_\mathcal{V}}\sum_{j=1}^{N_\mathcal{V}} p_{ij}=1
\right\}.
\label{Eq_feasible_set}
\end{equation}
The set $\mathcal P$ is a closed and bounded convex subset of 
$\mathbb R^{{N_\mathcal{V}}({N_\mathcal{V}}-1)}$, and hence compact. Rewrite the marginal distributions
\begin{equation}
    p_i^{\mathrm{out}}=\sum_{j=1}^{N_\mathcal{V}} p_{ij},
    \qquad
    p_j^{\mathrm{in}}=\sum_{i=1}^{N_\mathcal{V}} p_{ij}.
\end{equation}
Since the function $x\mapsto -x\ln x$ is continuous on $\left[0,1\right]$ and the marginals are linear mappings of ${\bm{p}}$, both $e({\bm{p}})$ and $r({\bm{p}})$ are continuous on $\mathcal{P}$. Whenever $e({\bm{p}})+r({\bm{p}})>0$, we have
\begin{equation}
    \alpha({\bm{p}})=\frac{e({\bm{p}})}{e({\bm{p}})+r({\bm{p}})}\in[0,1],
\end{equation}
thus $\alpha$ is continuous on $\mathcal P$ (with continuous extension at the boundary). Let us construct two uniformly distributed extreme network structures: a complete network and a unidirectional ring network, as shown in Fig.~\ref{Fig_TwoNetworks}.

\begin{figure}[htb!]
    \centering
    \includegraphics[width=1\linewidth]{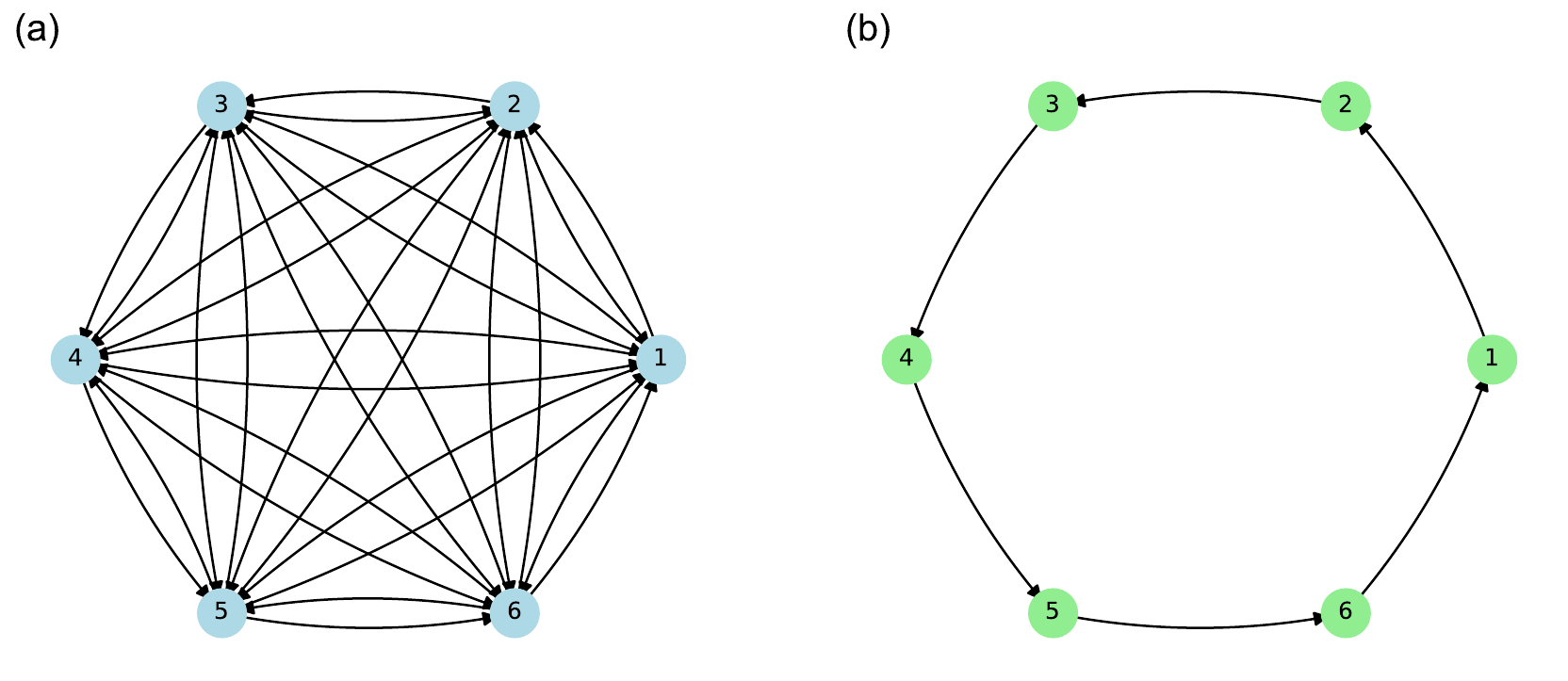}
    \caption{Complete network (a) and unidirectional ring network (b).}
    \label{Fig_TwoNetworks}
\end{figure}

\emph{(i) Complete network.}
Let $\bar{\bm{p}}=\{\bar{p}_{ij}: \forall i,j \in \mathcal{V}\}$, where
\begin{equation}
    \bar{p}_{ij}=
    \begin{cases}
        \dfrac{1}{N_\mathcal{V}({N_\mathcal{V}}-1)}, & i\neq j,\\
        0,& i=j.
    \end{cases}
\end{equation}

Then, we have
\begin{equation}
    e(\bar{\bm{p}})
    =\sum\limits_{i=1}^{N_\mathcal{V}}\sum\limits_{j=1}^{N_\mathcal{V}} \frac{1}{{N_\mathcal{V}}({N_\mathcal{V}}-1)}\ln\frac{{N_\mathcal{V}}({N_\mathcal{V}}-1)}{({N_\mathcal{V}}-1)^2}
    =\ln\frac{{N_\mathcal{V}}}{{N_\mathcal{V}}-1},
\end{equation}
\begin{equation}
    r(\bar{\bm{p}})
    =\sum\limits_{i=1}^{N_\mathcal{V}}\sum\limits_{j=1}^{N_\mathcal{V}} \frac{1}{{N_\mathcal{V}}({N_\mathcal{V}}-1)}\ln({N_\mathcal{V}}-1)^2
    =2\ln({N_\mathcal{V}}-1),
\end{equation}
and
\begin{equation}
    \alpha(\bar{\bm{p}})
    =\dfrac{\ln{N_\mathcal{V}}-\ln({{N_\mathcal{V}}-1})}{\ln{N_\mathcal{V}}+\ln({{N_\mathcal{V}}-1})},
\end{equation}
where $\alpha(\bar{\bm{p}})<1/\mathrm{e}$ for ${N_\mathcal{V}}\geq 3$.

\emph{(ii) Unidirectional ring network.} Let $\tilde{\bm{p}}=\{\tilde{p}_{ij}: \forall i,j \in \mathcal{V}\}$, where
\begin{equation}
    \tilde{p}_{ij}=
    \begin{cases}
        \dfrac{1}{N_\mathcal{V}}, & j=\bmod(i+1,N_\mathcal{V}),\\
        0,& \text{otherwise}.
    \end{cases}
\end{equation}

Then, we have
\begin{equation}
    e(\tilde{\bm{p}}) =\sum\limits_{i=1}^{N_\mathcal{V}}\sum\limits_{j=1}^{N_\mathcal{V}} \frac{1}{{N_\mathcal{V}}}\ln\frac{{N_\mathcal{V}}^2}{{N_\mathcal{V}}}
    =\ln {N_\mathcal{V}},
\end{equation}
\begin{equation}
    r(\tilde{\bm{p}})
    =\sum\limits_{i=1}^{N_\mathcal{V}}\sum\limits_{j=1}^{N_\mathcal{V}} \frac{1}{{N_\mathcal{V}}}\ln\frac{{N_\mathcal{V}}^2}{{N_\mathcal{V}}^2}
    =0,
\end{equation}
and
\begin{equation}
    \alpha(\tilde{\bm{p}})=1,
\end{equation}
indicating this network is overly-efficient and possesses deterministic structure.

\emph{(iii) Combining the two networks.} Since $\mathcal P$ is convex, for any $\lambda\in[0,1]$, let
\begin{equation}
    {\bm{p}}(\lambda)=(1-\lambda)\bar{\bm{p}}+\lambda\tilde{\bm{p}}\in\mathcal P.
\end{equation}
The function $\alpha\left({\bm{p}}(\lambda)\right)$ of $\lambda$ is continuous on $[0,1]$ and satisfies
\begin{equation}
    \alpha\left({\bm{p}}(0)\right)=\alpha\left(\bar{\bm{p}}\right)<\frac{1}{\mathrm{e}},
    \qquad
    \alpha\left({\bm{p}}(1)\right)=\alpha\left(\tilde{\bm{p}}\right)=1>\frac{1}{\mathrm{e}}.
\end{equation}
By the intermediate value theorem, there exists $\lambda^*\in(0,1)$ such that
\begin{equation}
   \alpha\left({\bm{p}}(\lambda^*)\right)=\frac{1}{\mathrm{e}}.
\end{equation}
Setting ${\bm{p}}^*={\bm{p}}(\lambda^*)$ completes the proof. Consequently, for weighted and directed networks with at least three nodes and no self-loops, there exists at least one joint distribution such that the resilience is optimal ($R=1/\mathrm{e}$).
\end{proof}

\begin{lemma}
\label{Lem_FixedDistribution}
Let $N_\mathcal{V} \geq 3$. In the set of all feasible network configurations $\mathcal{P}$ defined in Eq.~(\ref{Eq_feasible_set}), if the marginal distributions remain uniform, i.e.,
\begin{equation}
    p_i^{\mathrm{out}}=p_j^{\mathrm{in}}=\frac{1}{N_\mathcal{V}}, \quad \forall i,j \in \mathcal{V},
\end{equation}
then the redundancy $r$ and the ratio of order $\alpha$ can be analytically reduced to
\begin{equation}
    r = 2\ln N_\mathcal{V}-2e
\end{equation}
and
\begin{equation}
    \alpha = \dfrac{e}{2\ln N_\mathcal{V} -e}.
\end{equation}
Consequently, the optimal resilience condition $\alpha=1/\mathrm{e}$ is strictly equivalent to
\begin{equation}
    e=\frac{2\ln N_\mathcal{V}}{\mathrm{e}+1}.
\end{equation}
\end{lemma}

\begin{proof}
The resulting joint probabilities of Theorem~\ref{Thm_AtleastThreeNodes} are given by
\begin{equation}
    p_{ij}(\lambda)=
    \begin{cases}
        \dfrac{1+\lambda(N_\mathcal{V}-2)}{N_\mathcal{V}(N_\mathcal{V}-1)}, 
        & j=\bmod(i+1,N_\mathcal{V}),\\[6pt]
        \dfrac{1-\lambda}{N_\mathcal{V}(N_\mathcal{V}-1)}, 
        & j\neq i,\ \bmod(i+1,N_\mathcal{V}),\\[6pt]
        0, & i=j.
    \end{cases}
    \label{Eq_p_ij_lambda}
\end{equation}
For all $\lambda\in[0,1]$, the marginal distributions remain
uniform:
\begin{equation}
    p_i^{\mathrm{out}}(\lambda)=\frac{1}{N_\mathcal{V}},
    \qquad
    p_j^{\mathrm{in}}(\lambda)=\frac{1}{N_\mathcal{V}},
    \qquad
    \forall\, i,j.
\end{equation}
Hence, variations in $\lambda$ affect only the joint structure of flows, while the marginals retain maximal entropy. When $p_i^{\mathrm{out}}=p_j^{\mathrm{in}}=\dfrac{1}{N_\mathcal{V}}$, we have
\begin{equation}
\begin{aligned}
    r
    &=-\sum_{i=1}^{N_\mathcal{V}}\sum_{j=1}^{N_\mathcal{V}}
    p_{ij}\ln\left(\frac{p_{ij}}{p_i^{\mathrm{out}}}\right)-\sum_{i=1}^{N_\mathcal{V}}\sum_{j=1}^{N_\mathcal{V}}
    p_{ij}\ln\left(
    \frac{p_{ij}}{p_j^{\mathrm{in}}}\right) \\
    &=-2\sum_{i=1}^{N_\mathcal{V}}\sum_{j=1}^{N_\mathcal{V}}
    p_{ij}\ln\left(N_\mathcal{V} p_{ij}\right) \\
    &=-2\ln N_\mathcal{V}+2H \\
    &=2\ln N_\mathcal{V} - 2e.
\end{aligned}
\end{equation}
Consequently, the order parameter $\alpha$ takes the closed form
\begin{equation}
    \alpha
    =\frac{e}{e+r}
    =\frac{e}{2\ln N_\mathcal{V} - e}.
\end{equation}
The optimal condition $\alpha=1/\mathrm{e}$ is equivalent to
\begin{equation}
    e=\frac{2\ln N_\mathcal{V}}{\mathrm{e}+1}.
\end{equation}
\end{proof}

\begin{figure}[htb!]
    \centering
    \includegraphics[width=1\linewidth]{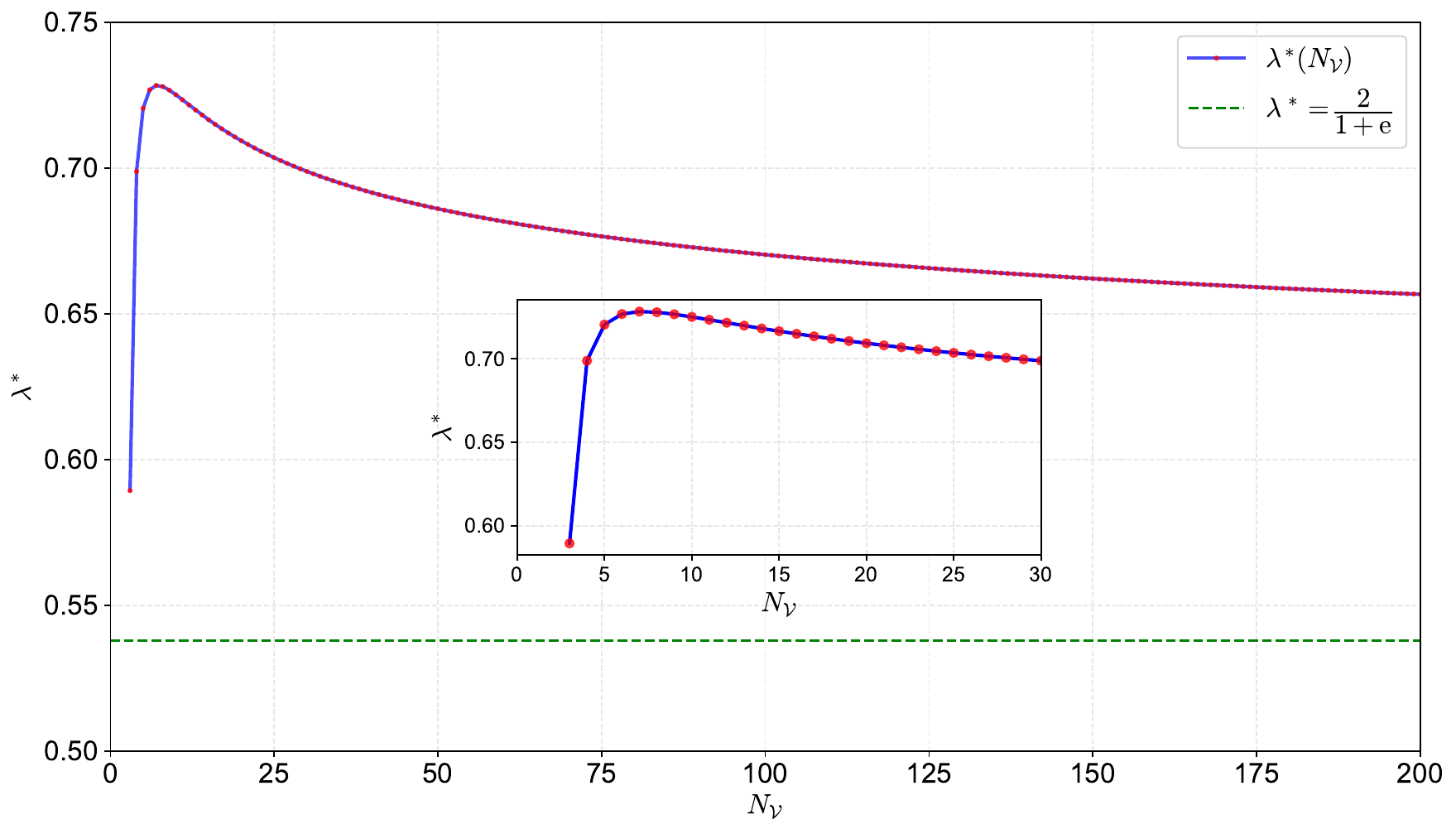}
    \caption{The relationship between $N_\mathcal{V}$ and the parameter of the optimal condition $\lambda^*$.}
    \label{Fig_Lambda_N_V}
\end{figure}

According to Lemma~\ref{Lem_FixedDistribution}, for the joint probabilities given by Eq.~(\ref{Eq_p_ij_lambda}), we have
\begin{equation}
\begin{aligned}
    e(\lambda)
    &=\sum_{i=1}^{N_\mathcal{V}}\sum_{j=1}^{N_\mathcal{V}}
    p_{ij}(\lambda)\ln\left(
    \frac{p_{ij}(\lambda)}{p_i^{\mathrm{out}}p_j^{\mathrm{in}}}
    \right) \\
    &=\sum_{i=1}^{N_\mathcal{V}}\sum_{j=1}^{N_\mathcal{V}} p_{ij}(\lambda)\ln\left(N_\mathcal{V}^2 p_{ij}(\lambda)\right) \\
    &=\frac{1+\lambda(N_\mathcal{V}-2)}{N_\mathcal{V}-1}
    \ln\left(
    \frac{N_\mathcal{V}\left(1+\lambda(N_\mathcal{V}-2)\right)}{N_\mathcal{V}-1}
    \right)
    +\frac{(N_\mathcal{V}-2)(1-\lambda)}{N_\mathcal{V}-1}
    \ln\left(
    \frac{N_\mathcal{V}(1-\lambda)}{N_\mathcal{V}-1}
    \right)
\end{aligned}
\end{equation}
and
\begin{equation}
\begin{aligned}
    r(\lambda)
    &=-\sum_{i=1}^{N_\mathcal{V}}\sum_{j=1}^{N_\mathcal{V}}
    p_{ij}(\lambda)\ln\left(\frac{p_{ij}(\lambda)}{p_i^{\mathrm{out}}}\right)-\sum_{i=1}^{N_\mathcal{V}}\sum_{j=1}^{N_\mathcal{V}}
    p_{ij}(\lambda)\ln\left(
    \frac{p_{ij}(\lambda)}{p_j^{\mathrm{in}}}\right) \\
    &=-2\sum_{i=1}^{N_\mathcal{V}}\sum_{j=1}^{N_\mathcal{V}}
    p_{ij}(\lambda)\ln\left(N_\mathcal{V} p_{ij}(\lambda)\right) \\
    &=2\ln N_\mathcal{V} - 2e(\lambda).
\end{aligned}
\end{equation}
Consequently, the order parameter $\alpha(\lambda)$ takes the closed form
\begin{equation}
    \alpha(\lambda)
    =\frac{e(\lambda)}{e(\lambda)+r(\lambda)}
    =\frac{e(\lambda)}{2\ln N_\mathcal{V} - e(\lambda)}.
\end{equation}
The optimal condition $\alpha(\lambda^*)=1/\mathrm{e}$ is equivalent to
\begin{equation}
    e(\lambda^*)=\frac{2\ln N_\mathcal{V}}{\mathrm{e}+1}.
\end{equation}
Substituting the explicit expression of $e(\lambda^*)$ yields the following scalar equation for
$\lambda^*$:
\begin{equation}
    \frac{1+\lambda^*(N_\mathcal{V}-2)}{N_\mathcal{V}-1}
    \ln\left(
    \frac{N_\mathcal{V}\left(1+\lambda^*(N_\mathcal{V}-2)\right)}{N_\mathcal{V}-1}
    \right)
    +\frac{(N_\mathcal{V}-2)(1-\lambda^*)}{N_\mathcal{V}-1}
    \ln\left(
    \frac{N_\mathcal{V}(1-\lambda^*)}{N_\mathcal{V}-1}
    \right)
    =\frac{2\ln N_\mathcal{V}}{\mathrm{e}+1},
\end{equation}
which has a unique solution $\lambda^*\in(0,1)$ for any $N_\mathcal{V}\ge3$. When $N_\mathcal{V}=7$, $\lambda^*$ reaches its maximum value (0.728). When $N_\mathcal{V} \to+\infty$, $\lambda^*$ converges to $\dfrac{2}{\mathrm{e}+1}>0.5$, as shown in Fig.~\ref{Fig_Lambda_N_V}. Consequently, for the interpolated family between a complete directed graph and a deterministic ring, the optimal configuration has an explicit characterization through a one-dimensional equation. The results reveal that the more nodes there are, the more the interpolating family needs to be biased towards efficient structures to achieve the optimal resilience.

\section{Case analysis: symmetric directed network}
\label{Section_Results}

Before delving into the detailed analytical derivations, we briefly outline the logical progression of this section. While Section~\ref{Section_Existence} proves the general existence of optimal resilience for weighted and directed networks with $N_\mathcal{V} \ge 3$ and no self-loops, extracting explicit scaling behaviors for arbitrary heterogeneous topologies is mathematically intractable. Therefore, to investigate the mechanics of the efficiency-redundancy trade-off, we construct a parameterized symmetric network model defined by three distinct flow types: adjacent forward links ($x$), adjacent backward links ($y$), and non-adjacent background cross-links ($z$).

Using this model, we analyze four representative boundary cases: (1) $x=y$ (symmetric forward and backward cycles), (2) $y=z$ (uniform background and backward flows), (3) $z=0$ (the complete absence of background cross-links), and (4) $y=0$ (purely forward primary flows with background redundancy). Studying these specific configurations allows us to demonstrate mathematically how different structural components govern the optimal state. More importantly, it enables us to derive explicit asymptotic scaling laws ($N_\mathcal{V} \to \infty$), revealing the absolute necessity of background redundancy ($z$) in sustaining optimal resilience as network size increases.

\subsection{Governing equations}

\begin{figure}[htb!]
    \centering
    \includegraphics[width=1\linewidth]{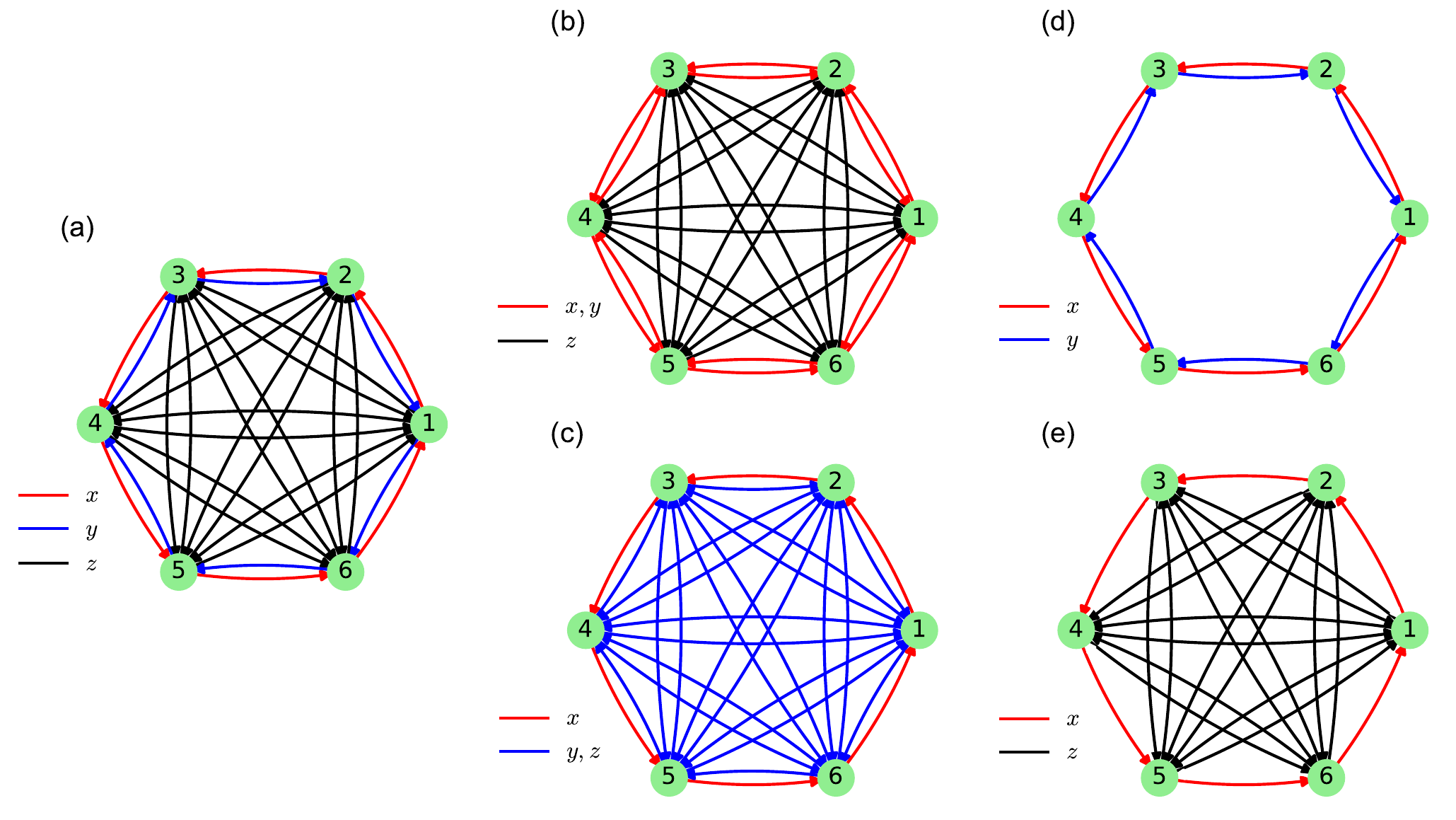}
    \caption{The symmetric directed network with parameters $(x,y,z)$ represented by Eq.~(\ref{Eq_Network_xyz}) (a) and four special cases: (b) $x=y$, (c) $y=z$, (d) $z=0$, and (e) $y=0$.}
    \label{Fig_Network_4SpecialCases}
\end{figure}

For a weighted and directed graph with $N_\mathcal{V}$ nodes and no self-loops ($p_{ii} = 0$ for all $i$), the joint probability matrix $\{p_{ij}\}$ is defined based on the given conditions. The matrix is an $N_\mathcal{V} \times N_\mathcal{V}$ matrix where the elements are arranged as follows: (1) The elements immediately above the diagonal (in a circular sense) are $x$, i.e., $p_{i,i+1} = x$ for $i = 1, \dots, N_\mathcal{V}-1$, and  $p_{N_\mathcal{V},1} = x$; (2) The elements immediately below the diagonal (in a circular sense) are $y$, i.e., $p_{i+1,i} = y$ for $i = 1, \dots, N_\mathcal{V}-1$, and  $p_{1,N_\mathcal{V}} = y$; and (3) All other off-diagonal elements are $z$. In other words, we have
\begin{equation}
p_{ij} = 
\begin{cases}
x, & \text{if} j \equiv i+1 \pmod{N_\mathcal{V}} {\text{~~(adjacent forward link)}}, \\
y, & \text{if } j \equiv i-1 \pmod{N_\mathcal{V}} {\text{~~(adjacent backward link)}}, \\
z, & \text{otherwise, for } i \neq j  {\text{~~(non-adjacent background cross-link)}}, \\
0, & \text{if } i = j  {\text{~~(diagonal)}}.
\end{cases}
\label{Eq_Network_xyz}
\end{equation}
The matrix can be represented as
\begin{equation}
\{p_{ij}\} = \begin{pmatrix}
0 & x & z & \cdots & z & y \\
y & 0 & x & \cdots & z & z \\
z & y & 0 & \cdots & z & z \\
\vdots & \vdots & \vdots & \ddots & \vdots & \vdots \\
z & z & z & \cdots & 0 & x \\
x & z & z & \cdots & y & 0
\end{pmatrix}.
\end{equation}
A schematic representation of this network structure is shown in Fig.~\ref{Fig_Network_4SpecialCases}(a).

The probability normalization condition ensures that the sum of all probabilities is 1. The number of $x$ elements is  $N_\mathcal{V}$, the number of $y$ elements is $N_\mathcal{V}$, and the number of $z$ elements is $N_\mathcal{V}(N_\mathcal{V} - 3)$. Thus, the probability normalization equation is
\begin{equation}
    N_\mathcal{V} x + N_\mathcal{V} y + N_\mathcal{V} (N_\mathcal{V} - 3) z = 1.
\end{equation}
Simplifying, we get
\begin{equation}
    x + y + (N_\mathcal{V} - 3) z = \frac{1}{N_\mathcal{V}}.
\end{equation}
Due to symmetry, all nodes have identical marginals:
\begin{equation}
    p_i^{\mathrm{out}} = p_j^{\mathrm{in}} = x + y + (N_\mathcal{V}-3)z = \frac{1}{N_\mathcal{V}},
\end{equation}
which is the same as the probability normalization condition (Lemma~\ref{Lem_FixedDistribution}).

Accordingly, we have
\begin{equation}
e = \sum_{i=1}^{N_{\mathcal{V}}}\sum_{j=1}^{N_{\mathcal{V}}} p_{ij} \ln \frac{p_{ij}}{p_{i}^{\mathrm{out}} p_{j}^{\mathrm{in}}}
= \sum_{i=1}^{N_{\mathcal{V}}}\sum_{j=1}^{N_{\mathcal{V}}} p_{ij} \ln p_{ij} + 2 \ln N_\mathcal{V},
\end{equation}
where
\begin{equation}
\sum_{i=1}^{N_{\mathcal{V}}}\sum_{j=1}^{N_{\mathcal{V}}} p_{ij} \ln p_{ij} = N_\mathcal{V} [ x \ln x + y \ln y + (N_\mathcal{V}-3) z \ln z ].
\end{equation}
Therefore, the efficiency $e$ can be expressed as
\begin{equation}
    e = N_\mathcal{V}[x \ln x + y \ln y + (N_\mathcal{V}-3)z \ln z] + 2\ln N_\mathcal{V}.
\end{equation}
For redundancy $r$, we have
\begin{equation}
   r
   = 2\ln N_\mathcal{V}-2e 
   = -2N_\mathcal{V} [ x \ln x + y \ln y + (N_\mathcal{V}-3) z \ln z ]-2\ln N_\mathcal{V}.
\end{equation}
Hence, we have
\begin{equation}
\begin{aligned}
   \alpha &= \frac{e}{2\ln N_\mathcal{V} - e}\\
    &= -\frac{2\ln N_\mathcal{V}}{N_\mathcal{V}[x \ln x + y \ln y + (N_\mathcal{V}-3)z \ln z]}-1.
\end{aligned}
\end{equation}
The resilience $R$ is maximized when $\alpha = 1/\mathrm{e}$, which occurs when
\begin{equation}
e = \frac{2\ln N_\mathcal{V}}{\mathrm{e}+1}.
\end{equation}
Furthermore, we have
\begin{equation}
    x \ln x + y \ln y + (N_\mathcal{V}-3)z \ln z
    = -\frac{2\mathrm{e}}{\mathrm{e}+1}\cdot\frac{\ln N_\mathcal{V}}{N_\mathcal{V}}.
\end{equation}
Finally, we obtain the following equations:
\begin{equation}
    \begin{cases}
       x + y + (N_\mathcal{V} - 3) z = \dfrac{1}{N_\mathcal{V}}, \\
       x \ln x + y \ln y + (N_\mathcal{V}-3)z \ln z = -\dfrac{2\mathrm{e}}{\mathrm{e}+1} \cdot \dfrac{\ln N_\mathcal{V}}{N_\mathcal{V}},
    \end{cases}
    \label{Eq_xyz}
\end{equation}
where $x,y,z\geq0$. Subsequently, we consider four cases, as presented in Fig.~\ref{Fig_Network_4SpecialCases}(b-e).

\subsection{The case of \texorpdfstring{$x=y$}{x=y}}
\label{Subsection_x_equal_y}

\subsubsection{Existence of solutions}

Assume $x = y$. Then Eqs.~(\ref{Eq_xyz}) reduce to
\begin{equation}
    \begin{cases}
       2x + (N_\mathcal{V} - 3) z = \dfrac{1}{N_\mathcal{V}}, \\
       2x \ln x + (N_\mathcal{V}-3)z \ln z = -\dfrac{2\mathrm{e}}{\mathrm{e}+1} \cdot \dfrac{\ln N_\mathcal{V}}{N_\mathcal{V}},
    \end{cases}
\end{equation}
where $x,z\geq0$. The normalization condition yields
\begin{equation}
    x = \dfrac{1}{2N_\mathcal{V}} - \dfrac{(N_\mathcal{V} - 3) z}{2}.
\label{Eq_x_equal_y_NormalCondition}
\end{equation}
Substituting this expression into the second equation gives
\begin{equation}
    2\left[\dfrac{1}{2N_\mathcal{V}} - \dfrac{(N_\mathcal{V} - 3) z}{2}\right] \ln\left[\dfrac{1}{2N_\mathcal{V}} - \dfrac{(N_\mathcal{V} - 3) z}{2}\right] + (N_\mathcal{V}-3)z \ln z = -\dfrac{2\mathrm{e}}{\mathrm{e}+1} \cdot \dfrac{\ln N_\mathcal{V}}{N_\mathcal{V}}.
\label{Eq_x_equal_y}
\end{equation}
The feasibility condition requires
\begin{equation}
    0\leq z \leq \frac{1}{N_\mathcal{V}(N_\mathcal{V} - 3)}.
\end{equation}
To analyze the solution properties of Eq.~(\ref{Eq_x_equal_y}), define
\begin{equation}
    f(z) = 2\left[\frac{1}{2N_\mathcal{V}} - \frac{(N_\mathcal{V} - 3)z}{2}\right] \ln\left[\frac{1}{2N_\mathcal{V}} - \frac{(N_\mathcal{V} - 3)z}{2}\right] + (N_\mathcal{V}-3)z \ln z + \frac{2\mathrm{e}}{\mathrm{e}+1} \cdot \frac{\ln N_\mathcal{V}}{N_\mathcal{V}}.
\end{equation}
Let $ a = N_\mathcal{V} - 3 $ (with $ a > 0 $) and define
\begin{equation}
u(z) = \frac{1}{2N_\mathcal{V}} - \frac{(N_\mathcal{V} - 3)z}{2}.
\end{equation}
Then
\begin{equation}
f(z) = 2u(z) \ln u(z) + a z \ln z + \frac{2\mathrm{e}}{\mathrm{e}+1} \cdot \frac{\ln N_\mathcal{V}}{N_\mathcal{V}}.
\end{equation}
The first derivative of $f(z)$ is
\begin{equation}
\begin{aligned}
f'(z) &= 2\left[u'(z) \ln u(z) + u(z) \cdot \frac{u'(z)}{u(z)}\right] + a\left[\ln z + z \cdot \frac{1}{z}\right] \\
&= 2\left[u'(z) \ln u(z) + u'(z)\right] + a(\ln z + 1) \\
&= 2u'(z) \left(\ln u(z) + 1\right) + a\left(\ln z + 1\right) \\
&= a\ln\left(\frac{z}{u(z)}\right).
\end{aligned}
\end{equation}
Hence, $f(z)$ is monotonically decreasing when $0<z<\dfrac{1}{N_\mathcal{V}(N_\mathcal{V}-1)}$ and monotonically increasing when $\dfrac{1}{N_\mathcal{V}(N_\mathcal{V}-1)}<z<\dfrac{1}{N_\mathcal{V}(N_\mathcal{V}-3)}$. The minimum value is attained at
\begin{equation}
    z=\dfrac{1}{N_\mathcal{V}(N_\mathcal{V}-1)},
\end{equation}
with
\begin{equation}
    f(z)_\mathrm{min}=f\left(\dfrac{1}{N_\mathcal{V}(N_\mathcal{V}-1)}\right)=\dfrac{1}{N_\mathcal{V}}\left(\dfrac{\mathrm{e}-1}{\mathrm{e}+1}\ln N_\mathcal{V}-\ln(N_\mathcal{V}-1)\right).
\end{equation}
For $N_\mathcal{V}>3$, $f(z)_\mathrm{min}<0$.

The second derivative is
\begin{equation}
f''(z) = a \cdot \frac{d}{dz} \left[\ln z - \ln u(z)\right]
= \frac{a}{z} + \frac{a^2}{2u(z)},
\end{equation}
which is non-negative on the feasible domain, indicating that $f(z)$ is convex.

\begin{figure}[htb!]
    \centering
    \includegraphics[width=1\linewidth]{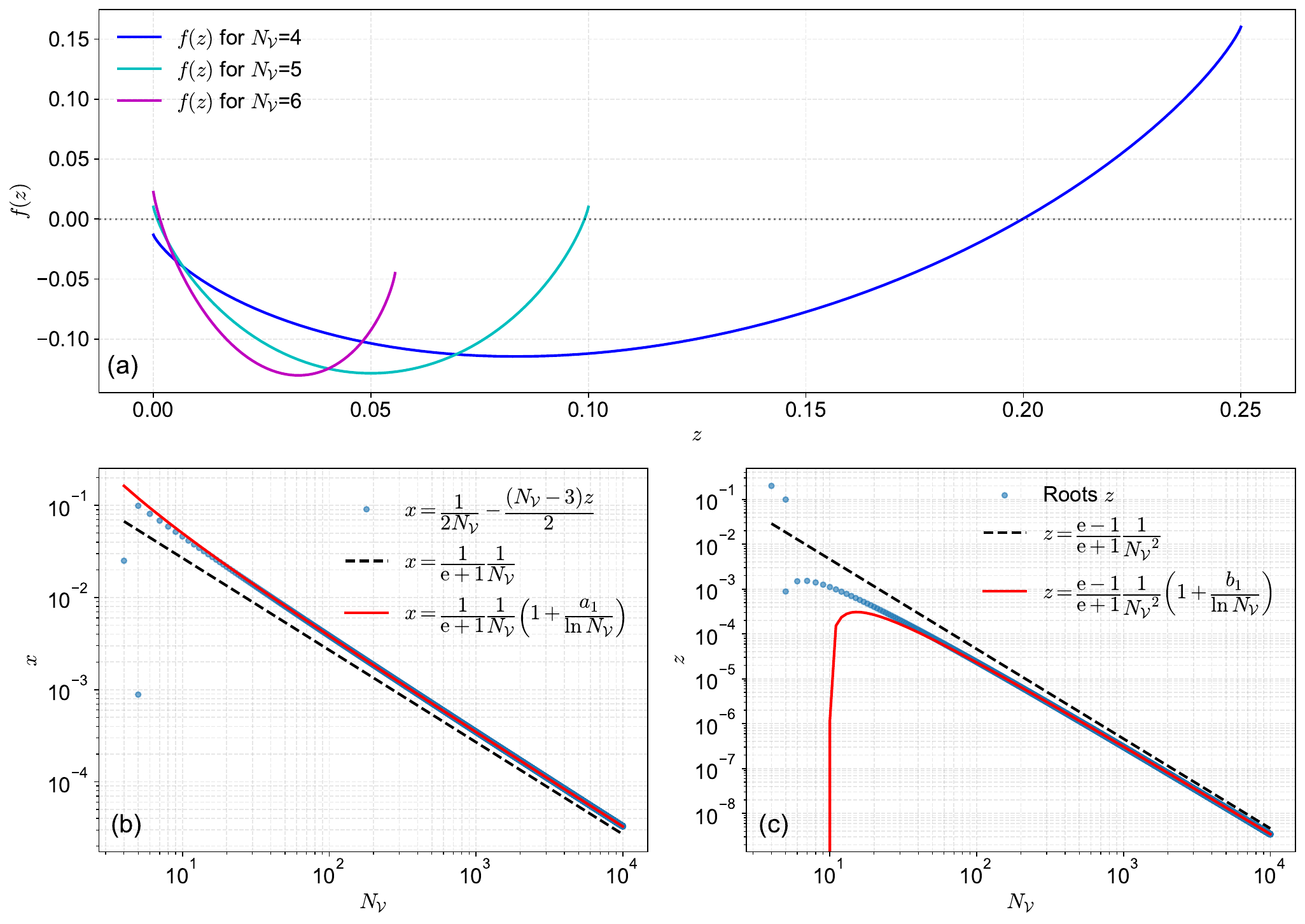}
    \caption{Existence of optimal resilience and asymptotic scaling behavior of link weights for the symmetric case $x=y$. \normalfont{(a) The governing function $f(z)$ illustrating the existence of roots for small network sizes ($N_\mathcal{V}=4,5,6$). (b) Asymptotic scaling of the adjacent link weight $x$ and (c) the non-adjacent link weight $z$ as a function of network size $N_\mathcal{V}$ in log-log scale. In panels (b) and (c), the scattered blue dots represent the exact numerical solutions (roots). The lines correspond to the analytical asymptotic approximations: the black dashed lines denote the first-order leading terms (scaling as $O(N_\mathcal{V}^{-1})$ for $x$ and $O(N_\mathcal{V}^{-2})$ for $z$), and the solid red lines represent the second-order approximations including logarithmic corrections, where parameters $a_1$ and $b_1$ are defined in Eq.~(\ref{Eq_x_equal_y_a1_b1}).}}
    \label{Fig_fz_equal_0_x_equal_y}
\end{figure}

At the boundary points, we have
\begin{equation}
    f(0) = -\frac{1}{N_\mathcal{V}} \ln(2N_\mathcal{V}) + \frac{2\mathrm{e}}{\mathrm{e}+1} \cdot \frac{\ln N_\mathcal{V}}{N_\mathcal{V}}
\end{equation}
and
\begin{equation}
    f(\dfrac{1}{N_\mathcal{V}(N_\mathcal{V}-3)})
    = -\frac{1}{N_\mathcal{V}} \ln(N_\mathcal{V}(N_\mathcal{V}-3)) + \frac{2\mathrm{e}}{\mathrm{e}+1} \cdot \frac{\ln N_\mathcal{V}}{N_\mathcal{V}}.
\end{equation}
Combining the convexity of $f(z)$ and the boundary values, we conclude that: (1) when $N_\mathcal{V}=4$, we have $f(0)<0$ and $f\left(\dfrac{1}{N_\mathcal{V}(N_\mathcal{V}-3)}\right)>0$, thus Eq.~(\ref{Eq_x_equal_y}) has a unique solution; (2) when $N_\mathcal{V}=5$, we have $f(0)>0$ and $f\left(\dfrac{1}{N_\mathcal{V}(N_\mathcal{V}-3)}\right)>0$, thus Eq.~(\ref{Eq_x_equal_y}) has two solutions; and (3) when $N_\mathcal{V}\geq6$, we have $f(0)>0$ and $f\left(\dfrac{1}{N_\mathcal{V}(N_\mathcal{V}-3)}\right)<0$, thus Eq.~(\ref{Eq_x_equal_y}) has a unique solution.
The above situations are illustrated in Fig.~\ref{Fig_fz_equal_0_x_equal_y}(a), which reveals the existence of solutions of the optimal resilience when $x=y$.

\subsubsection{Asymptotic behavior of solutions}

Furthermore, by numerically solving $f(z)=0$ for increasing
$N_\mathcal{V}$, we observe that when $N_\mathcal{V}$ becomes sufficiently large (e.g., $N_\mathcal{V}>10^2$), the relationship between $N_\mathcal{V}$ and the corresponding root $z$ forms an approximately straight line in the double-logarithmic coordinate system, which suggests an asymptotic scaling with network size. To describe the asymptotic behavior of the root pair $(x,z)$ for large network size $N_{\mathcal{V}}$, we introduce scaled variables
\begin{equation}
    A = 2N_{\mathcal{V}}\,x,\qquad 
    B = (N_{\mathcal{V}}-3)\,N_{\mathcal{V}}\,z,
\end{equation}
such that Eq.~(\ref{Eq_x_equal_y_NormalCondition}) is equivalent to
\begin{equation}
    A+B=1.
\end{equation}
The entropy balance equation
\begin{equation}
    2x\ln x+(N_{\mathcal{V}}-3)z\ln z
    = -\frac{2\mathrm{e}}{\mathrm{e}+1}\cdot\frac{\ln N_{\mathcal{V}}}{N_{\mathcal{V}}}
\end{equation}
can then be recast in terms of $A$ and $B$ as
\begin{equation}
    A\ln A + B\ln B - A\ln 2 - B\ln(N_{\mathcal{V}}-3)
    = \left(\frac{1-\mathrm{e}}{\mathrm{e}+1}\right)\ln N_{\mathcal{V}}.
\end{equation}
As $N_{\mathcal{V}}\to\infty$, we make the asymptotic ansatz
\begin{equation}
    A = A_{0}\left(1 + \frac{a_{1}}{\ln N_{\mathcal{V}}} + o\left(\frac{1}{\ln N_{\mathcal{V}}}\right)\right),
    \qquad
    B = B_{0}\left(1 + \frac{b_{1}}{\ln N_{\mathcal{V}}} + o\left(\frac{1}{\ln N_{\mathcal{V}}}\right)\right),
\end{equation}
where $A_{0}$ and $B_{0}$ are the leading-order constants solving the $\ln N_{\mathcal{V}}$-independent balance:
\begin{equation}
    A_{0} + B_{0} = 1,
    \qquad
    A_{0} = \frac{2}{\mathrm{e}+1},\qquad
    B_{0} = \frac{\mathrm{e}-1}{\mathrm{e}+1}.
\end{equation}
Solving for the first-order logarithmic corrections gives
\begin{equation}
    a_{1} =
    \frac{1-\mathrm{e}}{2}b_1,
    \qquad
    b_{1} = \ln(\mathrm{e}-1) - \frac{\mathrm{e}+1}{\mathrm{e}-1}\ln(\mathrm{e}+1),
\label{Eq_x_equal_y_a1_b1}
\end{equation}
thus leading to the following asymptotic expansions for $x$ and $z$:
\begin{equation}
\begin{aligned}
    x 
    &= \frac{A}{2N_{\mathcal{V}}}
    = \frac{1}{\mathrm{e}+1}\cdot\frac{1}{N_{\mathcal{V}}}
    \left[
        1 + \frac{a_{1}}{\ln N_{\mathcal{V}}}
        + O\left(\frac{1}{(\ln N_{\mathcal{V}})^2}\right)
    \right], \\
    z 
    &= \frac{B}{N_{\mathcal{V}}(N_{\mathcal{V}}-3)}
    = \frac{\mathrm{e}-1}{\mathrm{e}+1}\cdot\frac{1}{{N_\mathcal{V}}^2}
    \left[
        1 + \frac{b_{1}}{\ln N_{\mathcal{V}}}
        + O\left(\frac{1}{(\ln N_{\mathcal{V}})^2}\right)
    \right].
\end{aligned}
\end{equation}

This result explains why, despite the apparent asymptotic scaling decay of $z$, the quantity $x$ scales linearly with $N_{\mathcal{V}}^{-1}$, appearing as a straight line with slope $-1$ in a log-log plot. The logarithmic correction enters only at subleading order and becomes numerically negligible for moderate values of $N_\mathcal{V}$, accounting for the agreement between theory and numerical solutions observed in Figs.~\ref{Fig_fz_equal_0_x_equal_y}(b-c). Hence, when $x=y$, the link weights $x$, $y$, and $z$ that make the network resilience optimal exhibit an asymptotic scaling as $N_\mathcal{V}$ is large.

\subsection{The case of \texorpdfstring{$y=z$}{y=z}}

\subsubsection{Existence of solutions}

Assume $y = z$. Then Eqs.~(\ref{Eq_xyz}) reduce to
\begin{equation}
    \begin{cases}
       x + (N_\mathcal{V} - 2)z = \dfrac{1}{N_\mathcal{V}}, \\
       x \ln x + (N_\mathcal{V}-2)z \ln z
       = -\dfrac{2\mathrm{e}}{\mathrm{e}+1} \cdot \dfrac{\ln N_\mathcal{V}}{N_\mathcal{V}},
    \end{cases}
\end{equation}
where $x,z\geq0$. The normalization condition yields
\begin{equation}
    x = \frac{1}{N_\mathcal{V}} - (N_\mathcal{V}-2)z.
\label{Eq_y_equal_z_NormalCondition}
\end{equation}
Substituting this expression into the second equation gives
\begin{equation}
    (N_\mathcal{V}-2)z \ln z
    +\left[\frac{1}{N_\mathcal{V}}-(N_\mathcal{V}-2)z\right]
    \ln\left[\frac{1}{N_\mathcal{V}}-(N_\mathcal{V}-2)z\right]= -\dfrac{2\mathrm{e}}{\mathrm{e}+1} \cdot \dfrac{\ln N_\mathcal{V}}{N_\mathcal{V}}.
    \label{Eq_y_equal_z}
\end{equation}
The feasibility condition requires
\begin{equation}
    0 \leq z \leq \frac{1}{N_\mathcal{V}(N_\mathcal{V}-2)}.
\end{equation}
To analyze the solution properties of Eq.~(\ref{Eq_y_equal_z}), we define the function $f(z)$ analogously. Following a similar derivative analysis to Section~\ref{Subsection_x_equal_y}, it is straightforward to verify that $f(z)$ is strictly convex on the feasible domain ($f''(z) > 0$) and attains its unique minimum at $z = \dfrac{1}{N_\mathcal{V}(N_\mathcal{V}-1)}$, where $f(z)_\mathrm{min} < 0$ for $N_\mathcal{V} > 3$. Evaluating the boundary points yields $f(0) > 0$ and $f\left(\dfrac{1}{N_\mathcal{V}(N_\mathcal{V}-2)}\right) < 0$ for $N_\mathcal{V} \geq 4$. Combining the convexity of $f(z)$ with these boundary values, we conclude that the equation has a unique optimal solution for $N_\mathcal{V} \geq 4$. This confirms the existence of optimal resilience when $y=z$.

\subsubsection{Asymptotic behavior of solutions}

Furthermore, numerical roots suggest an asymptotic scaling with network size. Applying the identical asymptotic ansatz methodology from Section~\ref{Subsection_x_equal_y}, we introduce the scaled variables $A = N_{\mathcal{V}}x$ and $B = (N_{\mathcal{V}}-2)N_{\mathcal{V}}z$ with $A+B=1$. By balancing the logarithmic terms as $N_{\mathcal{V}} \to \infty$, the leading-order constants remain $A_0 = \dfrac{2}{\mathrm{e}+1}$ and $B_0 = \dfrac{\mathrm{e}-1}{\mathrm{e}+1}$. The first-order logarithmic corrections are found to be
\begin{equation}
    a_{1} = \frac{1-\mathrm{e}}{2}b_1, \qquad b_{1} = \ln(\mathrm{e}-1) + \frac{2\ln 2}{\mathrm{e}-1} - \frac{\mathrm{e}+1}{\mathrm{e}-1}\ln(\mathrm{e}+1),
    \label{Eq_y_equal_z_a1_b1}
\end{equation}
thus leading to the following explicit asymptotic expansions for $x$ and $z$:
\begin{equation}
\begin{aligned}
    x 
    &= \frac{A}{N_{\mathcal{V}}}
    = \frac{2}{\mathrm{e}+1}\cdot\frac{1}{N_\mathcal{V}}
    \left[
        1 + \frac{a_{1}}{\ln N_{\mathcal{V}}}
        + O\left(\frac{1}{(\ln N_{\mathcal{V}})^2}\right)
    \right], \\
    z 
    &= \frac{B}{N_{\mathcal{V}}(N_{\mathcal{V}}-2)}
    = \frac{\mathrm{e}-1}{\mathrm{e}+1}\cdot\frac{1}{{N_\mathcal{V}}^2}
    \left[
        1 + \frac{b_{1}}{\ln N_{\mathcal{V}}}
        + O\left(\frac{1}{(\ln N_{\mathcal{V}})^2}\right)
    \right].
\end{aligned}
\end{equation}

\begin{figure}[htb!]
    \centering
    \includegraphics[width=1\linewidth]{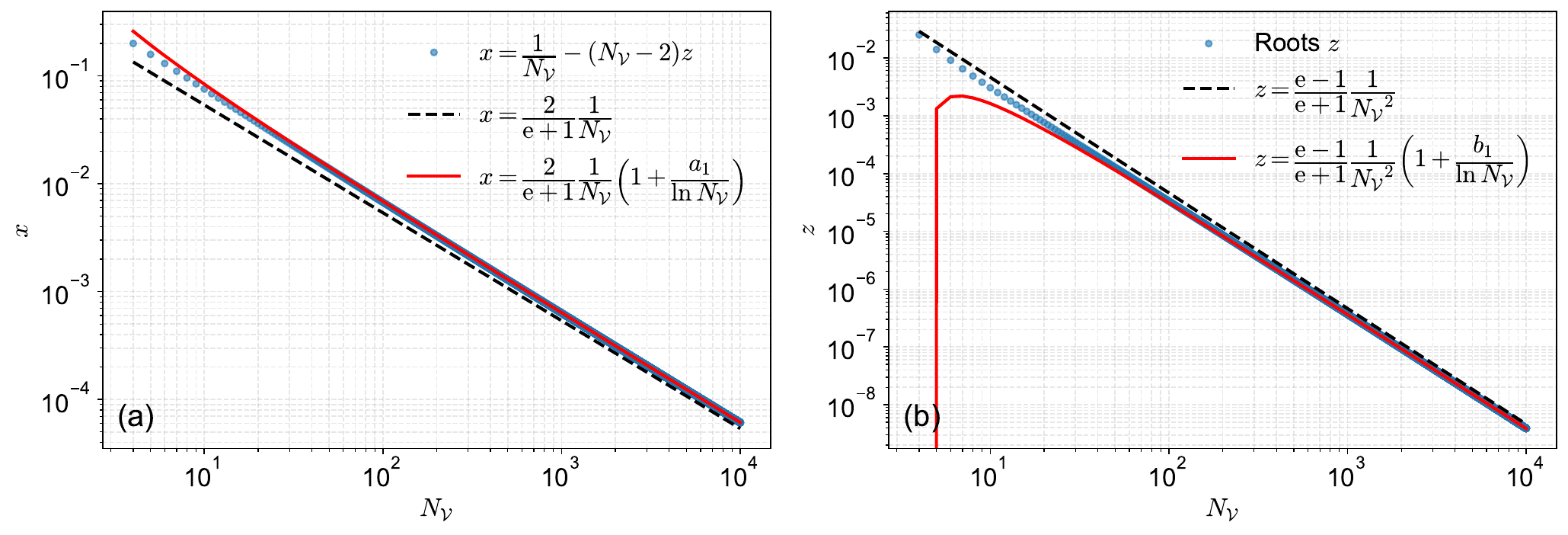}
    \caption{Asymptotic scaling behavior of link weights for the symmetric case $y=z$. \normalfont{(a) Asymptotic scaling of the adjacent link weight $x$ and (b) the non-adjacent link weight $z$ as a function of network size $N_\mathcal{V}$ in log-log scale. In panels (a) and (b), the scattered blue dots represent the exact numerical solutions (roots). The lines correspond to the analytical asymptotic approximations: the black dashed lines denote the first-order leading terms (scaling as $O(N_\mathcal{V}^{-1})$ for $x$ and $O(N_\mathcal{V}^{-2})$ for $z$), and the solid red lines represent the second-order approximations including logarithmic corrections, where parameters $a_1$ and $b_1$ are defined in Eq.~(\ref{Eq_y_equal_z_a1_b1}).}}
    \label{Fig_fz_equal_0_y_equal_z}
\end{figure}

This result is similar to the result when $x=y$, where the quantity $x$ remains tightly constrained to a straight line with slope $-1$ in the double-logarithmic representation. The logarithmic correction enters only at subleading order and becomes numerically negligible for moderate values of $N_\mathcal{V}$, accounting for the agreement between theory and numerical solutions observed in Figs.~\ref{Fig_fz_equal_0_y_equal_z}(a-b). Hence, when $y=z$, the link weights $x$, $y$, and $z$ that make the network resilience optimal exhibit an asymptotic scaling as $N_\mathcal{V}$ is large.

\subsection{The case of \texorpdfstring{$z=0$}{z=0}}

\begin{figure}[htb!]
    \centering
    \includegraphics[width=1\linewidth]{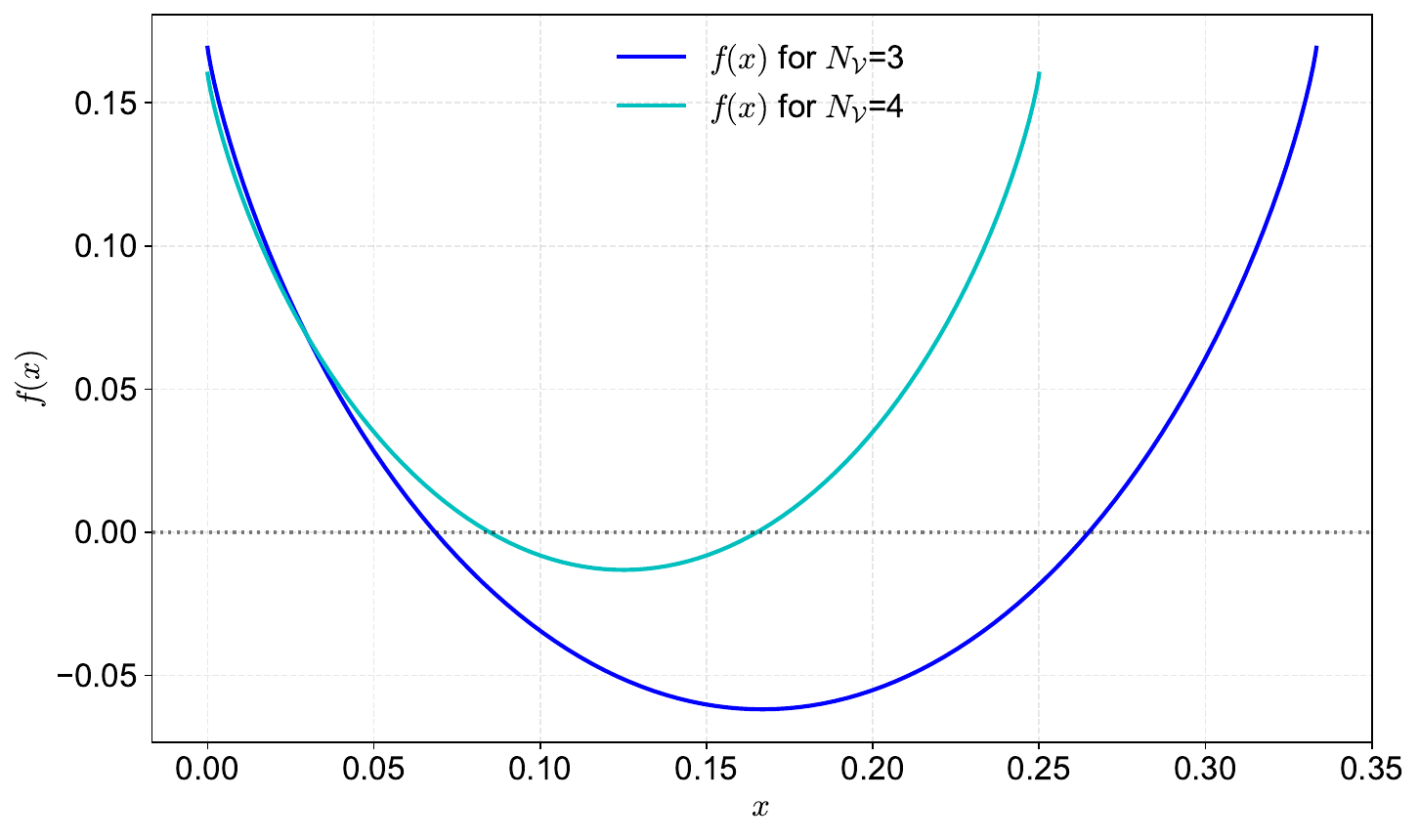}
    \caption{Existence of optimal resilience for the symmetric case $z=0$. \normalfont{The governing function $f(x)$ has solutions only when network sizes $N_\mathcal{V}=3$ or $N_\mathcal{V}=4$.}}
    \label{Fig_fx_equal_0_z_equal_0}
\end{figure}

Assume $z=0$. Eqs.~(\ref{Eq_xyz}) reduce to
\begin{equation}
    \begin{cases}
       x + y = \dfrac{1}{N_\mathcal{V}}, \\
       x \ln x + y \ln y = -\dfrac{2\mathrm{e}}{\mathrm{e}+1} \cdot \dfrac{\ln N_\mathcal{V}}{N_\mathcal{V}},
    \end{cases}
\end{equation}
where $x,y \ge 0$. The normalization condition yields
\begin{equation}
    y = \frac{1}{N_\mathcal{V}} - x.
\end{equation}
Substituting this expression into the second equation gives
\begin{equation}
    x \ln x + \left(\frac{1}{N_\mathcal{V}} - x\right)\ln\left(\frac{1}{N_\mathcal{V}} - x\right)
    = -\dfrac{2\mathrm{e}}{\mathrm{e}+1} \cdot \dfrac{\ln N_\mathcal{V}}{N_\mathcal{V}}.
    \label{Eq_z_equal_0}
\end{equation}
The feasible condition requires
\begin{equation}
    0 \le x \le \frac{1}{N_\mathcal{V}}.
\end{equation}
To analyze the solution structure of Eq.~(\ref{Eq_z_equal_0}), we define
\begin{equation}
    f(x) = x \ln x + \left(\frac{1}{N_\mathcal{V}} - x\right)\ln\left(\frac{1}{N_\mathcal{V}} - x\right)
    + \dfrac{2\mathrm{e}}{\mathrm{e}+1} \cdot \dfrac{\ln N_\mathcal{V}}{N_\mathcal{V}}.
\end{equation}
Through a similar calculus procedure, $f(x)$ is found to be strictly convex ($f''(x) > 0$) on the feasible domain, reaching its minimum at $x = \dfrac{1}{2N_\mathcal{V}}$ with $f(x)_\mathrm{min} = \dfrac{1}{N_\mathcal{V}} \left[\dfrac{\mathrm{e}-1}{\mathrm{e}+1}\ln N_\mathcal{V}-\ln 2\right]$.
Evaluating the boundaries yields identical positive values: $f(0) = f\left(\dfrac{1}{N_\mathcal{V}}\right) > 0$. Consequently, roots only exist when the minimum value is strictly negative. Mathematical evaluation reveals that $f(x)_\mathrm{min} < 0$ exclusively for $N_\mathcal{V}=3$ or $N_\mathcal{V}=4$. For $N_\mathcal{V} > 4$, the minimum becomes positive, indicating no solutions exist. The above situations are illustrated in Fig.~\ref{Fig_fx_equal_0_z_equal_0}, which reveals that there exists the optimal resilience for networks with $N_\mathcal{V}=3$ or $N_\mathcal{V}=4$ when $z=0$.

Physically, this mathematical impossibility holds a structural implication. When $z=0$, the network is completely devoid of background cross-links, reducing to a ring-like topology driven solely by adjacent cycles. For very small networks ($N_\mathcal{V} \le 4$), this simple structure can still accidentally satisfy the entropy balance. However, as the network scales up ($N_\mathcal{V} > 4$), the system becomes overwhelmingly deterministic and overly efficient. Without the dispersed background cross-links ($z$) to provide alternative flow pathways, this purely ring-like network inherently cannot generate sufficient structural redundancy (conditional entropy) to counterbalance its high efficiency, making it structurally impossible to reach the $\alpha = 1/\mathrm{e}$ optimal threshold.

\subsection{The case of \texorpdfstring{$y=0$}{y=0}}

\subsubsection{Existence of solutions}

\begin{figure}[htb!]
    \centering
    \includegraphics[width=1\linewidth]{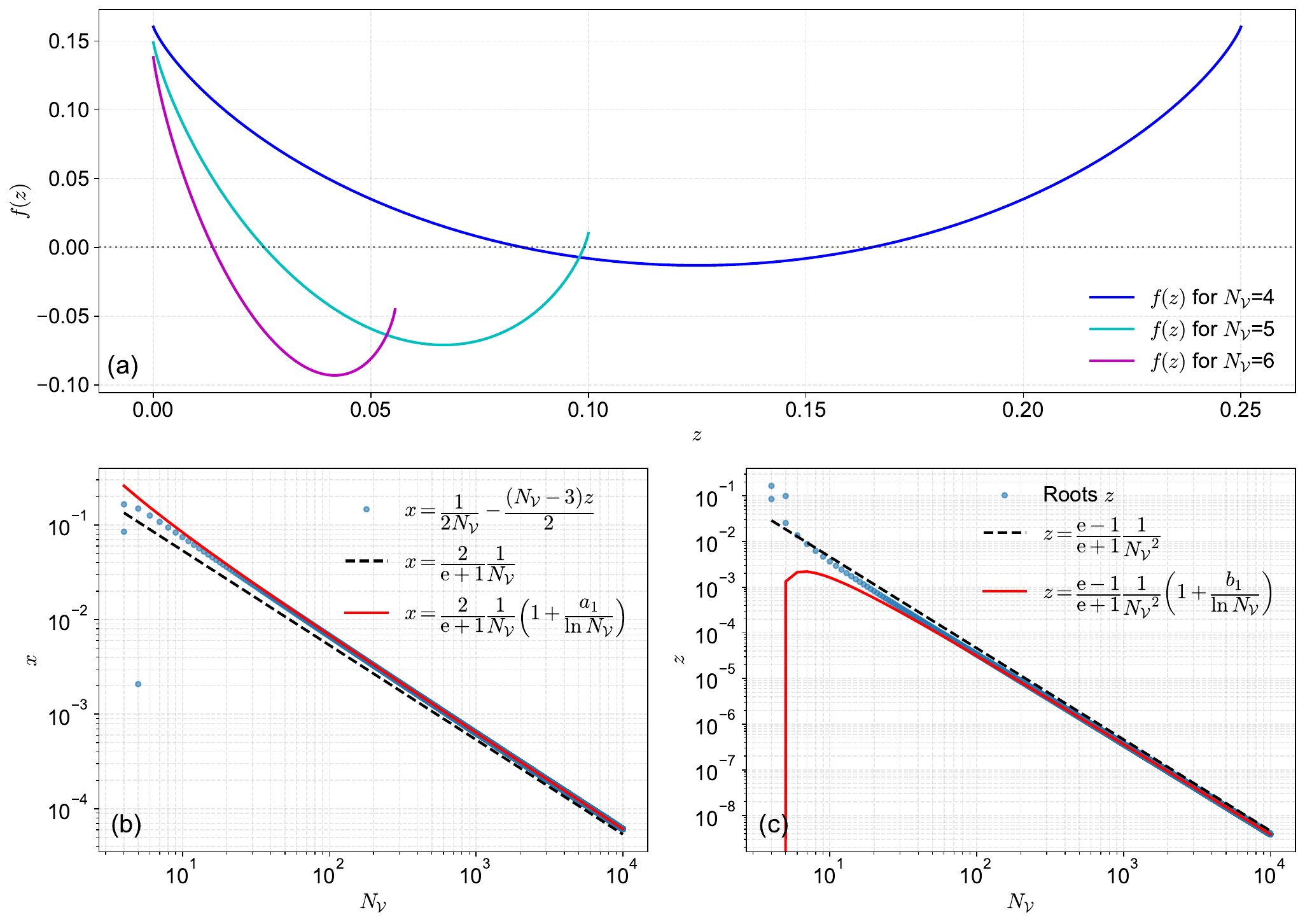}
    \caption{Existence of optimal resilience and asymptotic scaling behavior of link weights for the symmetric case $y=0$. \normalfont{(a) The governing function $f(z)$ illustrating the existence of roots for small network sizes ($N_\mathcal{V}=4,5,6$). (b) Asymptotic scaling of the adjacent link weight $x$ and (c) the non-adjacent link weight $z$ as a function of network size $N_\mathcal{V}$ in log-log scale. In panels (b) and (c), the scattered blue dots represent the exact numerical solutions (roots). The lines correspond to the analytical asymptotic approximations: the black dashed lines denote the first-order leading terms (scaling as $O(N_\mathcal{V}^{-1})$ for $x$ and $O(N_\mathcal{V}^{-2})$ for $z$), and the solid red lines represent the second-order approximations including logarithmic corrections, where parameters $a_1$ and $b_1$ are defined in Eq.~(\ref{Eq_y_equal_0_a1_b1}).}}
    \label{Fig_fz_equal_0_y_equal_0}
\end{figure}

Assume $y=0$. Then Eqs.~(\ref{Eq_xyz}) reduce to
\begin{equation}
    \begin{cases}
       x + (N_\mathcal{V} - 3) z = \dfrac{1}{N_\mathcal{V}}, \\
       x \ln x + (N_\mathcal{V}-3)z \ln z
       = -\dfrac{2\mathrm{e}}{\mathrm{e}+1} \cdot \dfrac{\ln N_\mathcal{V}}{N_\mathcal{V}},
    \end{cases}
\end{equation}
where $x,z \ge 0$. The normalization condition yields
\begin{equation}
    x = \dfrac{1}{N_\mathcal{V}} - (N_\mathcal{V} - 3) z.
\label{Eq_z_equal_0_NormalCondition}
\end{equation}
Substituting this expression into the second equation gives
\begin{equation}
\left[\dfrac{1}{N_\mathcal{V}} - (N_\mathcal{V} - 3) z\right]
\ln\left[\dfrac{1}{N_\mathcal{V}} - (N_\mathcal{V} - 3) z\right]
+ (N_\mathcal{V}-3)z \ln z
= -\dfrac{2\mathrm{e}}{\mathrm{e}+1} \cdot \dfrac{\ln N_\mathcal{V}}{N_\mathcal{V}}.
\label{Eq_y_equal_0}
\end{equation}
The feasible condition requires
\begin{equation}
    0 \le z \le \frac{1}{N_\mathcal{V}(N_\mathcal{V}-3)}.
\end{equation}
To analyze the solution properties of Eq.~(\ref{Eq_y_equal_0}), we define
\begin{equation}
    f(z)
    =
    \left[\frac{1}{N_\mathcal{V}} - (N_\mathcal{V} - 3) z\right]
    \ln\left[\frac{1}{N_\mathcal{V}} - (N_\mathcal{V} - 3) z\right]
    + (N_\mathcal{V}-3)z \ln z
    + \frac{2\mathrm{e}}{\mathrm{e}+1} \cdot \frac{\ln N_\mathcal{V}}{N_\mathcal{V}}
\end{equation}
and find it to be strictly convex. The minimum is attained at $z = \dfrac{1}{N_\mathcal{V}(N_\mathcal{V}-2)}$, where $f(z)_\mathrm{min} < 0$ for $N_\mathcal{V} > 3$. By evaluating the boundary values at $z=0$ and the upper feasible limit, and combining this with the convexity property, we determine that: (1) when $4 \le N_\mathcal{V} \le 5$, the function crosses zero twice, yielding two solutions; and (2) when $N_\mathcal{V} > 5$, the boundaries dictate a unique solution. The above situations are illustrated in Fig.~\ref{Fig_fz_equal_0_y_equal_0}(a), which reveals the existence of solutions of the optimal resilience when $y=0$.

\subsubsection{Asymptotic behavior of solutions}

By utilizing the scaled variables $A = N_{\mathcal{V}}x$ and $B = (N_{\mathcal{V}}-3)N_{\mathcal{V}}z$ and applying the asymptotic ansatz as $N_{\mathcal{V}} \to \infty$, we bypass the intermediate entropy balance steps to directly obtain the first-order logarithmic corrections:
\begin{equation}
    a_{1} = \frac{1-\mathrm{e}}{2}b_1, \qquad b_{1} = \ln(\mathrm{e}-1) + \frac{2\ln 2}{\mathrm{e}-1} - \frac{\mathrm{e}+1}{\mathrm{e}-1}\ln(\mathrm{e}+1),
    \label{Eq_y_equal_0_a1_b1}
\end{equation}
thus leading to the following asymptotic expansions for $x$ and $z$:
\begin{equation}
\begin{aligned}
    x 
    &= \frac{A}{N_{\mathcal{V}}}
    = \frac{2}{\mathrm{e}+1}\cdot\frac{1}{N_\mathcal{V}}
    \left[
        1 + \frac{a_{1}}{\ln N_{\mathcal{V}}}
        + O\left(\frac{1}{(\ln N_{\mathcal{V}})^2}\right)
    \right], \\
    z 
    &= \frac{B}{N_{\mathcal{V}}(N_{\mathcal{V}}-3)}
    = \frac{\mathrm{e}-1}{\mathrm{e}+1}\cdot\frac{1}{{N_\mathcal{V}}^2}
    \left[
        1 + \frac{b_{1}}{\ln N_{\mathcal{V}}}
        + O\left(\frac{1}{(\ln N_{\mathcal{V}})^2}\right)
    \right].
\end{aligned}
\end{equation}

This result is similar to the results when $x=y$ and $y=z$, where the quantity $x$ remains tightly constrained to a straight line with slope $-1$ in the double-logarithmic representation. The logarithmic correction enters only at subleading order and becomes numerically negligible for moderate values of $N_\mathcal{V}$, accounting for the agreement between theory and numerical solutions observed in Figs.~\ref{Fig_fz_equal_0_y_equal_0}(b-c). Hence, when $y=0$, the link weights $x$, $y$, and $z$ that make the network resilience optimal exhibit an asymptotic scaling as $N_\mathcal{V}$ is large.

\begin{table}[htb!]
    \centering
    \caption{Logical structure of the paper: General theoretical foundations and asymptotic scaling behaviors of symmetric network cases.}
    \label{Tab_Summary_Cases}
    \renewcommand{\arraystretch}{1.3}
    \resizebox{\textwidth}{!}{
    \begin{tabular}{cllll}
        \toprule
        Scope / Case & Network configuration & Network size ($N_\mathcal{V}$) & Existence of optimal resilience & Main asymptotic scaling ($N_\mathcal{V}\to\infty$) \\
        \midrule
        \multicolumn{5}{l}{\textbf{Part I: General theoretical foundations}} \\
        \midrule
        Theorem~\ref{Thm_TwoNodes} & Two-node networks & $N_\mathcal{V} = 2$ & Unattainable (Deterministic) & N/A \\
        Theorem~\ref{Thm_AtleastThreeNodes} & Generalized directed networks & $N_\mathcal{V} \ge 3$ & At least one optimal configuration & N/A \\
        \midrule
        \multicolumn{5}{l}{\textbf{Part II: Specialized symmetric network constructions (Asymptotic analysis)}} \\
        \midrule
        \multirow{3}{*}{(a)} & \multirow{3}{*}{$x=y$} 
        & $N_\mathcal{V} = 4$ & Unique solution & \multirow{3}{*}{$x \simeq \dfrac{1}{\mathrm{e}+1}N_\mathcal{V}^{-1}$,~~$z \simeq \dfrac{\mathrm{e}-1}{\mathrm{e}+1}N_\mathcal{V}^{-2}$} \\
        & & $N_\mathcal{V} = 5$ & Two solutions & \\
        & & $N_\mathcal{V} \ge 6$ & Unique solution & \\
        \midrule
        (b) & $y=z$ 
        & $N_\mathcal{V} \ge 4$ & Unique solution & $x \simeq \dfrac{2}{\mathrm{e}+1}N_\mathcal{V}^{-1}$,~~$z \simeq \dfrac{\mathrm{e}-1}{\mathrm{e}+1}N_\mathcal{V}^{-2}$ \\
        \midrule
        \multirow{2}{*}{(c)} & \multirow{2}{*}{$z=0$}
        & $3\le N_\mathcal{V}\le 4$ & Two solutions & \multirow{2}{*}{N/A (No solution for $N_\mathcal{V} > 4$)} \\
        & & $N_\mathcal{V} > 4$ & No solution & \\
        \midrule
        \multirow{2}{*}{(d)} & \multirow{2}{*}{$y=0$} 
        & $4 \le N_\mathcal{V} \le 5$ & Two solutions & \multirow{2}{*}{$x \simeq \dfrac{2}{\mathrm{e}+1}N_\mathcal{V}^{-1}$,~~$z \simeq \dfrac{\mathrm{e}-1}{\mathrm{e}+1}N_\mathcal{V}^{-2}$} \\
        & & $N_\mathcal{V} > 5$ & Unique solution & \\
        \bottomrule
    \end{tabular}}
\begin{flushleft}
\footnotesize
Notes: Part I establishes the foundational attainability of $\alpha = 1/\mathrm{e}$ across general feasible probability spaces. Part II details the explicit asymptotic scaling within parameterized symmetric models. For symmetric models, all configurations are analyzed for $N_\mathcal{V} \ge 3$. However, cases (a), (b), and (d) require $N_\mathcal{V} \ge 4$ to allow for the existence of non-adjacent links ($z \neq 0$). For $N_\mathcal{V}=3$, the network structurally simplifies to the $z=0$ case.
\end{flushleft}
\end{table}

\section{Conclusion and discussion}
\label{Section_Conclusion}

This study provides a foundational theoretical investigation of entropy-based optimal resilience in weighted and directed networks with no self-loops. By analyzing the resilience metric within Ulanowicz's framework, we establish a general existence theorem for optimal resilience configurations in networks with at least three nodes, while demonstrating the structural impossibility of achieving this optimal state in two-node systems. These general results clarify the fundamental role of network scale and topological articulation in enabling the efficiency-redundancy trade-off.

To make the analytical derivation mathematically tractable, we further introduce a parameterized symmetric multi-link network model. From this specialized ansatz, we derive explicit governing equations for optimal flow allocations and obtain closed-form asymptotic scaling laws as the network size tends to infinity. The analytical results show that adjacent links scale inversely with network size ($O(N_\mathcal{V}^{-1})$), while background links exhibit a steeper quadratic decay ($O(N_\mathcal{V}^{-2})$) with specific logarithmic corrections. Furthermore, our numerical validations confirm that these asymptotic approximations converge rapidly, holding remarkably well even for medium-sized networks (e.g., $N_\mathcal{V} \sim 50$ to $100$). Mathematically, this distinct scaling behavior establishes a profound magnitude gap between link weights ($x \gg z$) as the network expands. Physically, this mathematical separation provides a structural mechanism for how the system balances efficiency and redundancy. The heavier adjacent links form a high-throughput primary backbone that ensures flow efficiency, whereas the extremely faint but numerous background cross-links provide a dispersed web of alternative pathways for structural redundancy. Consequently, the $\alpha = 1/\mathrm{e}$ optimal state inherently dictates that a network must spontaneously differentiate into primary functional channels and sparse backup routes.

While this study establishes a rigorous mathematical foundation, several limitations are explicitly noted to properly contextualize our findings. First, our general existence theorem is proven over the entire feasible probability space of directed networks. This implies that the topology is mathematically permitted to reconfigure. However, it does not guarantee that the $\alpha = 1/\mathrm{e}$ optimal state is attainable for every arbitrary, fixed network topology where zero-weight edges cannot be activated. Furthermore, to make the analytical derivations mathematically tractable, our specific asymptotic scaling laws rely on a symmetric network construction with uniform marginal distributions. Real-world systems exhibit pronounced structural heterogeneity, therefore the extent to which such topological heterogeneity affects the attainability and specific scaling of optimal resilience remains an important open question.

Second, our framework is purely static and structural. It does not incorporate dynamic perturbation processes, operational costs, node capacity constraints, or adaptive flow behaviors. Therefore, our derived governing equations currently provide fundamental theoretical insights rather than direct engineering tools. However, by explicitly integrating these theoretical resilience formulas as strict constraint equations or penalties within cost-aware objective functions, future operations research could leverage this framework to guide the structural design of human-made logistics or infrastructure networks.

Finally, our conclusions are intrinsically tied to Ulanowicz's specific entropy-based measure of resilience ($\alpha = 1/\mathrm{e}$). Future research could aim to relax these symmetric, static, and fixed-topology assumptions, thereby bridging the theoretical gap between idealized entropy optimization and the robust design of complex, heterogeneous real-world networks.

\section*{Acknowledgment}

This work was partly supported by the National Natural Science Foundation of China (72171083) and the Fundamental Research Funds for the Central Universities.

\section*{Declaration of competing interest}

The authors declare that they have no known competing financial interests or personal relationships that could have appeared to influence the work reported in this paper.

\section*{Data availability}

Data will be made available on request.

\clearpage

\appendix
\section*{Appendix}
\renewcommand{\thesubsection}{Appendix~\Alph{subsection}}
\setcounter{subsection}{0}

\subsection{Quantitative error analysis of asymptotic approximations}
\renewcommand{\thetable}{\Alph{subsection}.\arabic{table}}
\setcounter{table}{0}
\begin{table}[htb!]
    \centering
    \scriptsize
    \caption{Relative approximation errors between exact numerical roots, first-order approximations, and second-order asymptotic formulas for the symmetric case $x=y$.}
    \label{Tab_Error_Analysis_xy}
    \renewcommand{\arraystretch}{1.2}
    \newcolumntype{d}[1]{D{.}{.}{#1}}
    \begin{tabular*}{\textwidth}{@{\extracolsep{\fill}}c d{1.7} d{1.15} d{1.15} d{1.7} d{1.15} d{1.15} @{}}
        \toprule
        \multirow{2}{*}{$N_\mathcal{V}$} & \multicolumn{3}{c}{Adjacent link weight ($x$)} & \multicolumn{3}{c}{Background link weight ($z$)} \\
        \cmidrule(lr){2-4} \cmidrule(lr){5-7}
        & \multicolumn{1}{c}{Exact root} & \multicolumn{1}{c}{1st-order (Error)} & \multicolumn{1}{c}{2nd-order (Error)} & \multicolumn{1}{c}{Exact root} & \multicolumn{1}{c}{1st-order (Error)} & \multicolumn{1}{c}{2nd-order (Error)} \\
        \midrule
        10 & 4.6109\text{e}-02 & 2.6894\text{e}-02\,(41.67\%) & 4.9979\text{e}-02\,(8.39\%) & 1.1118\text{e}-03 & 4.6212\text{e}-03\,(315.66\%) & 4.1601\text{e}-06\,(99.63\%) \\
        20 & 2.1584\text{e}-02 & 1.3447\text{e}-02\,(37.70\%) & 2.2319\text{e}-02\,(3.40\%) & 4.0188\text{e}-04 & 1.1553\text{e}-03\,(187.47\%) & 2.6811\text{e}-04\,(33.29\%) \\
        30 & 1.3912\text{e}-02 & 8.9647\text{e}-03\,(35.56\%) & 1.4174\text{e}-02\,(1.89\%) & 2.0408\text{e}-04 & 5.1346\text{e}-04\,(151.60\%) & 1.6617\text{e}-04\,(18.58\%) \\
        50 & 8.0376\text{e}-03 & 5.3788\text{e}-03\,(33.08\%) & 8.0964\text{e}-03\,(0.73\%) & 8.3505\text{e}-05 & 1.8485\text{e}-04\,(121.36\%) & 7.6145\text{e}-05\,(8.81\%) \\
        80 & 4.8727\text{e}-03 & 3.3618\text{e}-03\,(31.01\%) & 4.8781\text{e}-03\,(0.11\%) & 3.5774\text{e}-05 & 7.2206\text{e}-05\,(101.84\%) & 3.4299\text{e}-05\,(4.12\%) \\
        100 & 3.8471\text{e}-03 & 2.6894\text{e}-03\,(30.09\%) & 3.8437\text{e}-03\,(0.09\%) & 2.3771\text{e}-05 & 4.6212\text{e}-05\,(94.40\%) & 2.3127\text{e}-05\,(2.71\%) \\
        200 & 1.8550\text{e}-03 & 1.3447\text{e}-03\,(27.51\%) & 1.8463\text{e}-03\,(0.46\%) & 6.5487\text{e}-06 & 1.1553\text{e}-05\,(76.42\%) & 6.5367\text{e}-06\,(0.18\%) \\
        500 & 7.1357\text{e}-04 & 5.3788\text{e}-04\,(24.62\%) & 7.0895\text{e}-04\,(0.65\%) & 1.1526\text{e}-06 & 1.8485\text{e}-06\,(60.37\%) & 1.1642\text{e}-06\,(1.00\%) \\
        1000 & 3.4825\text{e}-04 & 2.6894\text{e}-04\,(22.77\%) & 3.4589\text{e}-04\,(0.68\%) & 3.0442\text{e}-07 & 4.6212\text{e}-07\,(51.80\%) & 3.0822\text{e}-07\,(1.25\%) \\
        10000 & 3.2855\text{e}-05 & 2.6894\text{e}-05\,(18.14\%) & 3.2665\text{e}-05\,(0.58\%) & 3.4300\text{e}-09 & 4.6212\text{e}-09\,(34.73\%) & 3.4669\text{e}-09\,(1.08\%) \\
        \bottomrule
    \end{tabular*}
\begin{flushleft}
\footnotesize
Notes: The relative error is calculated as $|(\mathrm{Exact} - \mathrm{Asymptotic}) / \mathrm{Exact}| \times 100\%$. The 1st-order approximations denote the leading scaling terms ($O(N_\mathcal{V}^{-1})$ and $O(N_\mathcal{V}^{-2})$), while the 2nd-order approximations include the derived logarithmic corrections.
\end{flushleft}
\end{table}

\begin{table}[htb!]
    \centering
    \scriptsize
    \caption{Relative approximation errors between exact numerical roots, first-order approximations, and second-order asymptotic formulas for the symmetric case $y=z$.}
    \label{Tab_Error_Analysis_yz}
    \renewcommand{\arraystretch}{1.2}
    \newcolumntype{d}[1]{D{.}{.}{#1}}
    \begin{tabular*}{\textwidth}{@{\extracolsep{\fill}}c d{1.7} d{1.15} d{1.15} d{1.7} d{1.15} d{1.15} @{}}
        \toprule
        \multirow{2}{*}{$N_\mathcal{V}$} & \multicolumn{3}{c}{Adjacent link weight ($x$)} & \multicolumn{3}{c}{Background link weight ($z$)} \\
        \cmidrule(lr){2-4} \cmidrule(lr){5-7}
        & \multicolumn{1}{c}{Exact root} & \multicolumn{1}{c}{1st-order (Error)} & \multicolumn{1}{c}{2nd-order (Error)} & \multicolumn{1}{c}{Exact root} & \multicolumn{1}{c}{1st-order (Error)} & \multicolumn{1}{c}{2nd-order (Error)} \\
        \midrule
        10 & 7.5568\text{e}-02 & 5.3788\text{e}-02\,(28.82\%) & 8.3767\text{e}-02\,(10.85\%) & 3.0540\text{e}-03 & 4.6212\text{e}-03\,(51.31\%) & 1.6233\text{e}-03\,(46.85\%) \\
        20 & 3.6232\text{e}-02 & 2.6894\text{e}-02\,(25.77\%) & 3.8415\text{e}-02\,(6.02\%) & 7.6487\text{e}-04 & 1.1553\text{e}-03\,(51.05\%) & 5.7924\text{e}-04\,(24.27\%) \\
        30 & 2.3641\text{e}-02 & 1.7929\text{e}-02\,(24.16\%) & 2.4694\text{e}-02\,(4.46\%) & 3.4616\text{e}-04 & 5.1346\text{e}-04\,(48.33\%) & 2.8796\text{e}-04\,(16.81\%) \\
        50 & 1.3849\text{e}-02 & 1.0758\text{e}-02\,(22.32\%) & 1.4287\text{e}-02\,(3.16\%) & 1.2815\text{e}-04 & 1.8485\text{e}-04\,(44.24\%) & 1.1427\text{e}-04\,(10.84\%) \\
        80 & 8.4902\text{e}-03 & 6.7235\text{e}-03\,(20.81\%) & 8.6926\text{e}-03\,(2.38\%) & 5.1407\text{e}-05 & 7.2206\text{e}-05\,(40.46\%) & 4.7593\text{e}-05\,(7.42\%) \\
        100 & 6.7360\text{e}-03 & 5.3788\text{e}-03\,(20.15\%) & 6.8777\text{e}-03\,(2.10\%) & 3.3306\text{e}-05 & 4.6212\text{e}-05\,(38.75\%) & 3.1223\text{e}-05\,(6.26\%) \\
        200 & 3.2921\text{e}-03 & 2.6894\text{e}-03\,(18.31\%) & 3.3408\text{e}-03\,(1.48\%) & 8.6258\text{e}-06 & 1.1553\text{e}-05\,(33.94\%) & 8.2959\text{e}-06\,(3.82\%) \\
        500 & 1.2851\text{e}-03 & 1.0758\text{e}-03\,(16.29\%) & 1.2979\text{e}-03\,(1.00\%) & 1.4355\text{e}-06 & 1.8485\text{e}-06\,(28.77\%) & 1.4042\text{e}-06\,(2.18\%) \\
        1000 & 6.3294\text{e}-04 & 5.3788\text{e}-04\,(15.02\%) & 6.3781\text{e}-04\,(0.77\%) & 3.6779\text{e}-07 & 4.6212\text{e}-07\,(25.65\%) & 3.6219\text{e}-07\,(1.52\%) \\
        10000 & 6.1050\text{e}-05 & 5.3788\text{e}-05\,(11.89\%) & 6.1283\text{e}-05\,(0.38\%) & 3.8958\text{e}-09 & 4.6212\text{e}-09\,(18.62\%) & 3.8717\text{e}-09\,(0.62\%) \\
        \bottomrule
    \end{tabular*}
\begin{flushleft}
\footnotesize
Notes: The relative error is calculated as $|(\mathrm{Exact} - \mathrm{Asymptotic}) / \mathrm{Exact}| \times 100\%$. The 1st-order approximations denote the leading scaling terms ($O(N_\mathcal{V}^{-1})$ and $O(N_\mathcal{V}^{-2})$), while the 2nd-order approximations include the derived logarithmic corrections.
\end{flushleft}
\end{table}

\begin{table}[htb!]
    \centering
    \scriptsize
    \caption{Relative approximation errors between exact numerical roots, first-order approximations, and second-order asymptotic formulas for the symmetric case $y=0$.}
    \label{Tab_Error_Analysis_y0}
    \renewcommand{\arraystretch}{1.2}
    \newcolumntype{d}[1]{D{.}{.}{#1}}
    \begin{tabular*}{\textwidth}{@{\extracolsep{\fill}}c d{1.7} d{1.15} d{1.15} d{1.7} d{1.15} d{1.15} @{}}
        \toprule
        \multirow{2}{*}{$N_\mathcal{V}$} & \multicolumn{3}{c}{Adjacent link weight ($x$)} & \multicolumn{3}{c}{Background link weight ($z$)} \\
        \cmidrule(lr){2-4} \cmidrule(lr){5-7}
        & \multicolumn{1}{c}{Exact root} & \multicolumn{1}{c}{1st-order (Error)} & \multicolumn{1}{c}{2nd-order (Error)} & \multicolumn{1}{c}{Exact root} & \multicolumn{1}{c}{1st-order (Error)} & \multicolumn{1}{c}{2nd-order (Error)} \\
        \midrule
        10 & 7.4497\text{e}-02 & 5.3788\text{e}-02\,(27.80\%) & 8.3767\text{e}-02\,(12.44\%) & 3.6433\text{e}-03 & 4.6212\text{e}-03\,(26.84\%) & 1.6233\text{e}-03\,(55.44\%) \\
        20 & 3.6025\text{e}-02 & 2.6894\text{e}-02\,(25.35\%) & 3.8415\text{e}-02\,(6.64\%) & 8.2207\text{e}-04 & 1.1553\text{e}-03\,(40.53\%) & 5.7924\text{e}-04\,(29.54\%) \\
        30 & 2.3557\text{e}-02 & 1.7929\text{e}-02\,(23.89\%) & 2.4694\text{e}-02\,(4.83\%) & 3.6210\text{e}-04 & 5.1346\text{e}-04\,(41.80\%) & 2.8796\text{e}-04\,(20.47\%) \\
        50 & 1.3821\text{e}-02 & 1.0758\text{e}-02\,(22.16\%) & 1.4287\text{e}-02\,(3.37\%) & 1.3147\text{e}-04 & 1.8485\text{e}-04\,(40.60\%) & 1.1427\text{e}-04\,(13.09\%) \\
        80 & 8.4801\text{e}-03 & 6.7235\text{e}-03\,(20.71\%) & 8.6926\text{e}-03\,(2.51\%) & 5.2207\text{e}-05 & 7.2206\text{e}-05\,(38.31\%) & 4.7593\text{e}-05\,(8.84\%) \\
        100 & 6.7297\text{e}-03 & 5.3788\text{e}-03\,(20.07\%) & 6.8777\text{e}-03\,(2.20\%) & 3.3715\text{e}-05 & 4.6212\text{e}-05\,(37.07\%) & 3.1223\text{e}-05\,(7.39\%) \\
        200 & 3.2906\text{e}-03 & 2.6894\text{e}-03\,(18.27\%) & 3.3408\text{e}-03\,(1.52\%) & 8.6769\text{e}-06 & 1.1553\text{e}-05\,(33.15\%) & 8.2959\text{e}-06\,(4.39\%) \\
        500 & 1.2849\text{e}-03 & 1.0758\text{e}-03\,(16.28\%) & 1.2979\text{e}-03\,(1.01\%) & 1.4388\text{e}-06 & 1.8485\text{e}-06\,(28.47\%) & 1.4042\text{e}-06\,(2.41\%) \\
        1000 & 6.3289\text{e}-04 & 5.3788\text{e}-04\,(15.01\%) & 6.3781\text{e}-04\,(0.78\%) & 3.6821\text{e}-07 & 4.6212\text{e}-07\,(25.50\%) & 3.6219\text{e}-07\,(1.64\%) \\
        10000 & 6.1048\text{e}-05 & 5.3788\text{e}-05\,(11.89\%) & 6.1283\text{e}-05\,(0.38\%) & 3.8964\text{e}-09 & 4.6212\text{e}-09\,(18.60\%) & 3.8717\text{e}-09\,(0.63\%) \\
        \bottomrule
    \end{tabular*}
\begin{flushleft}
\footnotesize
Notes: The relative error is calculated as $|(\mathrm{Exact} - \mathrm{Asymptotic}) / \mathrm{Exact}| \times 100\%$. The 1st-order approximations denote the leading scaling terms ($O(N_\mathcal{V}^{-1})$ and $O(N_\mathcal{V}^{-2})$), while the 2nd-order approximations include the derived logarithmic corrections.
\end{flushleft}
\end{table}

\clearpage
\subsection{Numerical methods and reproducibility}
\label{Appendix_Numerical}

To ensure full reproducibility, the numerical roots of the nonlinear governing equations were obtained using a two-step Python/SciPy algorithm. First, for initialization, the feasible domain $(0, z_{\max})$ was uniformly discretized ($10,000$ to $20,000$ samples) to isolate intervals containing sign changes. Second, Brent's method (\textit{scipy.optimize.brentq}) was applied to each bracket with a solver tolerance set to the SciPy standard default precision ($\text{xtol} = 2 \times 10^{-12}$). This procedure was executed for network sizes ranging from $N_\mathcal{V} = 4$ to $10,000$. The logic is summarized in Algorithm~\ref{Alg_root_finding}, and the complete Python source code is provided in the Supplementary Material.

\begin{algorithm}[htb!]
\caption{Numerical root-finding and asymptotic scaling evaluation}
\label{Alg_root_finding}
\begin{algorithmic}[1]
\Require Network size range $N_\mathcal{V} \in [4, 10000]$, grid resolution $M$ (e.g., $10000$)
\Ensure Exact roots $(x^*, z^*)$ for each $N_\mathcal{V}$
\For{each $N_\mathcal{V}$ in range}
    \State Define the nonlinear function $f(z, N_\mathcal{V})$ and upper bound $z_{\max}$
    \State Generate $M$ evenly spaced test points $\{z_1, z_2, \dots, z_M\}$ in $(0, z_{\max})$
    \State Evaluate function values $\{f(z_1), f(z_2), \dots, f(z_M)\}$
    \State Initialize empty list \textit{Roots}
    \For{$i = 1$ \textbf{to} $M-1$}
        \If{$f(z_i) \cdot f(z_{i+1}) < 0$}
            \State Apply Brent's method on interval $[z_i, z_{i+1}]$ to find precise root $z^*$
            \State Append $z^*$ to \textit{Roots}
        \EndIf
    \EndFor
    \State Calculate corresponding primary link weight $x^*$ using normalization condition
\EndFor
\State Plot $x^*$ and $z^*$ against $N_\mathcal{V}$ in log-log scale to compare with analytical expansions
\end{algorithmic}
\end{algorithm}

\clearpage
\bibliographystyle{elsarticle-num}
\bibliography{Bib1,Bib2,BibCHN,BibITN,BibRCE,BibRobustNet}

\clearpage

\end{document}